\documentclass[final]{elsart}

\usepackage[dvips]{graphicx}

\usepackage{amssymb}
\usepackage{amsmath}

\begin{document}

\begin{frontmatter}

\title{On the possibility of primary identification of individual cosmic ray showers}

\author[UNAM]{A. D. Supanitsky\thanksref{Tandar}\corauthref{cor}},
\author[UNAM]{G. Medina-Tanco},
\author[Tandar]{A. Etchegoyen\thanksref{CONICET}}
\thanks[CONICET]{Member of Carrera del Investigador Cient\'ifico, CONICET, Argentina.}
\address[UNAM]{Instituto de Ciencias Nucleares, UNAM, Circuito Exteriror S/N, Ciudad Universitaria,
M\'exico D. F. 04510, M\'exico.}
\address[Tandar]{Departamento de F\'isica, Comisi\'on Nacional de Energ\'ia At\'omica, Av. Gral. Paz 1499, 
Buenos Aires, Argentina.}
\corauth[cor]{Corresponding author. Present Address: Instituto de Ciencias Nucleares, UNAM, Circuito 
Exteriror S/N, Ciudad Universitaria, M\'exico D. F. 04510, M\'exico. E-mail: supanitsky@nucleares.unam.mx.}

\begin{abstract}
The transition between the Galactic and extragalactic cosmic ray components could take place either
in the region of the spectrum known as the second knee or in the ankle. There are several models of
the transition but it is not possible to confirm or even rule out any of them from the flux measurement
alone. Therefore, the measurement of the composition as a function of primary energy will play a fundamental 
role for the understanding of this phenomenon. 

In this work we study the possibility of primary identification in an event by event basis in the ankle
region, around $E = 10^{18}$ eV. We consider as case study the enhancements of the Pierre Auger Southern 
Observatory, which are under construction in Malag\"ue, Province of Mendoza, Argentina. We use a non-parametric 
technique to estimate the density functions, from Monte Carlo data, corresponding to different combination of mass 
sensitive parameters and type of primaries. These estimates are used to obtain the classification probability of 
protons and iron nuclei for the different combination of parameters considered. We find that, after considering 
all relevant fluctuations, the maximum classification probability obtained combining surface and fluorescence 
detectors parameters is of order of $90\%$. 
\end{abstract}

\end{frontmatter}

\section{Introduction}
\label{Int}

The cosmic ray energy spectrum presents three main features
observed by several experiments, the knee, observed at around $3-5
\times 10^{15}$ eV \cite{Kampert:04,Aglietta:04,Antoni:05}, the
second knee, observed at $4 \times 10^{17}$ eV
\cite{Nagano:84,Abu:01,Pravdin:03,HiRes:04} and the ankle. There
is evidence of a fourth feature situated at the end of the
spectrum, the so-called GZK suppression \cite{GZKAuger,GZKHiRes},
which would be originated by the interaction of the ultra-high
energy protons with the photons of the cosmic microwave background
radiation \cite{Greisen:66,Zatsepin:66}. For the case of heavier
nuclei a similar effect is expected because of their interaction
with photons from the infrared and microwave backgrounds
\cite{Allard:05a}.

The origin of the second knee is still unclear, it has been
interpreted as the end of the efficiency of the acceleration in
Galactic supernova remnant shock waves, a change in the diffusion
regime in our galaxy \cite{Hoerandel:03,Candia:02} or even the
transition between the Galactic and extragalactic components of
the cosmic rays \cite{Berezinski:04}.

The ankle is a broader feature, it has been observed by Fly's Eye
\cite{Abu:01}, Haverah Park \cite{Ave:01}, Yakutsk
\cite{Pravdin:03}, HiRes \cite{HiRes:04} and Auger \cite{GZKAuger}
in Hybrid mode at approximately the same energy, $\sim 3\times
10^{18}$ eV. AGASA also observed the ankle but at a higher energy,
$\sim 10^{19}$ eV \cite{Takeda:03}. The origin of the ankle is
also unknown, it can be interpreted as the transition between the
Galactic and extragalactic components \cite{Allard:05} or the
result of pair production by extragalactic protons after the
interaction with photons of the cosmic microwave background
radiation during propagation \cite{Berezinski:04}.

There are three main models of the Galactic-extragalactic
transition: (i) the mixed composition model \cite{Allard:05}, in
which the extragalactic sources inject a spectrum of masses of the
form of the corresponding to the low energy Galactic cosmic rays
and for which the transition take place in the ankle, (ii) the dip
model \cite{Berezinski:04}, in which the ankle is originated by
pair production of extragalactic protons that interact with the
photons of the cosmic microwave background radiation, in this
scenario the transition is given at the second knee and (iii) the
ankle model, a two-component transition from Galactic iron nuclei
to extragalactic protons at the ankle energy \cite{Wibig:05}.

In order to rule out, or even confirm any of those models, additional information
is necessary besides the energy spectrum shape and absolute intensity. Detailed
measurements of the composition as a function of energy would be extremely valuable
to break the present degeneracy among competing models for the Galactic-extragalactic
transition \cite{MedinaTanco:07,GMT_EMA:06}. Furthermore, this kind of information could
help to determine what are the highest energy accelerators in the Galaxy and
provide indicatives of the kind and level of magnetohydrodynamic turbulence present
in the intergalactic medium traversed by the lowest energy cosmic ray particles
\cite{CinziaICRC:07}.

Several experiments have measured the cosmic ray composition in
the region where the transition takes place. Nevertheless,
large discrepancies exist between different experiments and experimental
techniques \cite{Dova:05}. One of the main reasons behind the plurality of
sometimes contradicting results is that the composition is determined by
comparing experimental data against numerical shower simulations. These
simulations include models for the relevant hadronic interactions which are
extrapolations, over several orders of magnitude in center of mass energy, of 
accelerator data to cosmic ray energies. No doubt, this is a source of considerable 
uncertainties which are confirmed, to a certain extent, by the fact that there are 
experimental evidences of a deficit in muon content of simulated showers with respect 
to real data \cite{Engel:07}.

Several multi-parametric techniques for the mass identification of individual showers 
have been studied and discussed in the literature. The most popular include non-parametric 
density estimates \cite{Antoni:07,Chilingarian:89,Glasmacher:99} and neural networks 
\cite{Antoni:07,Tiba:05}. In particular, in this work we use a non-parametric density estimate
technique of Gaussian kernel superposition, improved by adaptive choice of the smoothing parameters, 
in order to estimate the identification probability of individual nuclei, under the assumption of 
a binary proton and iron sample. The corresponding formalism, developed here, is general 
and applies to a sample of any number components. However, even if the application is formally 
straightforward, a wide range of theoretical and experimental uncertainties renders, at present, 
the extension to more than two components dubious. 
  
As case study, the mass sensitive parameters that will be available from the 
enhancements of the Pierre Auger Southern Observatory are considered. Auger, in 
its original design is able to measure cosmic rays of energies above 
$3\times10^{18}$ eV with the surface array and $\lesssim 10^{18}$ eV in hybrid 
mode. The enhancements AMIGA (Auger Muons and Infill for the Ground Array) 
\cite{Etchegoyen:07} and HEAT (High Elevation Auger Telescopes) \cite{Klages:07}, 
will extend the energy range down to $10^{17}$ eV, encompassing the second knee 
and ankle region where the Galactic-extragalactic transition takes place. 

AMIGA will consist of 85 pairs of Cherenkov detectors and muon counters
of 30 m$^2$ plastic scintillators buried at $\sim 2.5$ m of depth.
These pairs constitute the AMIGA infills, which are bounded by two hexagons
of 5.9 and 23.5 km$^2$ corresponding to arrays of 433 m and 750 m spacing, 
respectively. The energies at which the AMIGA arrays attain full detection 
efficiency independently of the primary mass, are $\sim 10^{17}$ eV and 
$\sim 10^{17.6}$ eV for the 433 m and 750 m arrays respectively \cite{Medina:06}. 
On the other hand, HEAT consists of three additional telescopes with elevation 
angle ranging from $30^\circ$ to $58^\circ$ and located next to the fluorescence 
telescope building at Coihueco. They will be used in combination with the 
existing $3^\circ$ to $30^\circ$ elevation angle telescopes at that site, as 
well as in hybrid mode in conjunction with the AMIGA infills.

The paper is organized as follow: in section \ref{clastech} we introduce the classification 
technique used for the subsequent analyses together with an analytical example of application 
of this technique. In section \ref{sim} we describe the full Monte Carlo simulations used in 
the calculation of the classification probabilities. In section \ref{clprop} we present the 
non-parametric methods used in the calculation of the classification probabilities and the 
results obtained for different combination of surface and fluorescence mass sensitive parameters. 
In section \ref{compo} we estimate the uncertainty in the determination of the composition of a 
sample obtained by using the classification technique introduced earlier. Finally, the discussion 
and conclusions are presented in section \ref{conc}.

\section{Classification technique}
\label{clastech}

The event by event classification consists in determining the type of nucleus
that originated a given event. This is usually done comparing the experimental data 
with simulated data. For this purpose non-parametric techniques like Bayes classifiers 
are very useful tools \cite{Antoni:07,Chilingarian:89}.

From the experimental data or even from simulations, several observable parameters, 
chosen such that they are very sensitive to the chemical composition of the primary, 
can be obtained. Let $\vec{x}$ be a $d$-dimensional vector composed by the mass sensitive 
parameters considered, $S_{cl}=\{ A_1, \ldots, A_L \}$ the set of classes in which the
the events will be classified ($S_{cl}=\{Pr, Fe\}$ in this work) and $P(\vec{x}\ |A_i)$ 
the conditional density function for the class $A_i$. Therefore, the probability of $A_i$ 
given $\vec{x}$ can be obtained by using the Bayes theorem \cite{Bayes:58},
\begin{equation}
P(A_i\ |\vec{x}) = \frac{P(\vec{x}\ |A_i)\ p(A_i)}{\sum^L_{j=1} P(\vec{x}\ |A_j)\ p(A_j)},
\label{Bayes}
\end{equation}
where $p(A_i)$ gives the prior knowledge about the relative abundances of each class. 
In absence of any prior knowledge the prior probability distribution is assumed to be 
uniform,
\begin{equation}
p(A_i) = \frac{1}{L}\ \ \ \ \textrm{such that} \ \ \ \ \sum^L_{i=1} p(A_j) = 1.
\label{Prior}
\end{equation}

The classification of the event is given by the class with maximum probability, 
\begin{equation}
A^*(\vec{x}) = \underset{A \in S_{cl}}{\operatorname{argmax}} \left[ P(A|\vec{x}) \right],
\label{CLRule}
\end{equation}
where $A^*(\vec{x})$ is the class assigned to the event with parameters $\vec{x}$. 

In this work, we consider the classification into proton and iron nuclei, 
$S_{cl}=\{ Pr,\ Fe \}$. Therefore, in this particular case, if $P(Pr | \vec{x}) \geq 1/2$
the event with parameters $\vec{x}$ is classified as proton, otherwise iron.

Therefore, the distribution functions $P(\vec{x}\ |Pr)$ and $P(\vec{x}\ |Fe)$ are required 
to classify the experimental data into proton and iron nuclei. Estimators of these distribution 
functions are obtained by using a non-parametric method of superposition of Gaussian kernels  
using the data obtained from the detailed simulation of the showers, including the response of 
the detectors and taking into account the effects introduced by the reconstruction methods
(see section \ref{clprop}).

A simplified one-dimensional analytical example is introduced in the following paragraphs in order to 
better understand the results obtained considering parameter spaces of more than one dimension and 
using non-parametric techniques to estimate the distribution functions from Monte Carlo data. This 
particular example presents similar features to those found in the complete and more sophisticated 
analysis.

Let us consider the gamma distribution,
\begin{equation}
f(x ;\alpha, \sigma) = \frac{\alpha}{\sigma \Gamma(\alpha)} \left( \frac{\alpha\ x}{\sigma}  \right)^{\alpha-1}%
\exp\left( -\frac{\alpha\ x}{\sigma} \right) 
\label{Gamma}
\end{equation}
where $\alpha$ and $\sigma$ are parameters. It is easy to show that the mean value and the variance of
this distribution are given by $E[ x ] = \sigma$ and $Var[x] = \sigma^2 /\alpha$, respectively. 

Two classes, $S_{cl}=\{ a,b \}$, are considered. The corresponding distribution functions are 
$P(x\ | a)=f(x;\alpha, \sigma_a)$ and $P(x\ |b)=f(x;\alpha, \sigma_b)$, where the parameter $\alpha$ is the 
same for both in order to simplify the calculations. Therefore, assuming a uniform distribution 
for the prior probabilities, the probability of $a$ given $x$ is, 
\begin{equation}
P(a\ |x) = \frac{P(x\ | a)}{P(x\ | a)+P(x\ | b)} =% 
\frac{f(x;\alpha, \sigma_a)}{f(x;\alpha, \sigma_a)+f(x;\alpha, \sigma_b)}.
\label{Paq}
\end{equation}

If $x$ is a random variable distributed according to $P(x\ | a)$ or $P(x\ | b)$,
\begin{equation}
p_a = \frac{f(x;\alpha, \sigma_a)}{f(x;\alpha, \sigma_a)+f(x;\alpha, \sigma_b)},
\label{pa}
\end{equation}
is also a random variable. Therefore, the distribution functions of $p_a$ for the cases
in which $x$ is distributed according to $f(x;\alpha, \sigma_a)$ and $f(x;\alpha, \sigma_b)$
are,
\begin{equation}
\label{V}
P_s(p_a)=\left\{ 
\begin{array}{ll}
   f(x(p_a);\alpha, \sigma_s) \left| {\mathop{\displaystyle \frac{dx}{dp_a}(p_a) }} \right| &  \textrm{if}\ p_a \in I_p \\
                                  % &                  \\
   0                          & \textrm{if}\ p_a \notin I_p  
\end{array}  \right..
\end{equation}
where $s=a, b$, $I_p=\left[ 1/(1+(\sigma_a/\sigma_b \right)^\alpha),1)$ if $\sigma_a > \sigma_b$ and
$I_p=\left( 0,1/(1+(\sigma_a/\sigma_b)^\alpha) \right]$ if $\sigma_a < \sigma_b$ and $p_a$ can be obtained 
inverting Eq. (\ref{pa}), 
\begin{equation}
x(p_a) = \frac{\sigma_b \sigma_a}{\sigma_b-\sigma_a}\ %
\ln\left[ \left( \frac{\sigma_a}{\sigma_b} \right)^\alpha \left(\frac{1}{p_a}-1 \right)  \right].
\label{qpa}
\end{equation}

The distribution functions of $p_a$ depend on how separated, relative to their width, are 
$f(x;\alpha, \sigma_a)$ and $f(x;\alpha, \sigma_b)$. A measure of this separation is given by
the parameter,
\begin{equation}
\eta = \frac{|E[x_a]-E[x_b]|}{\sqrt{Var[x_a]+Var[x_b]}}=% 
\sqrt{\alpha}\ \frac{|\sigma_a-\sigma_b|}{\sqrt{\sigma_a^2+\sigma_b^2}}, 
\label{eta}
\end{equation}
which is larger for distributions that are more separated (relative to their width).  

\begin{figure}[!bt]
\begin{center}
\includegraphics[width=6.8cm]{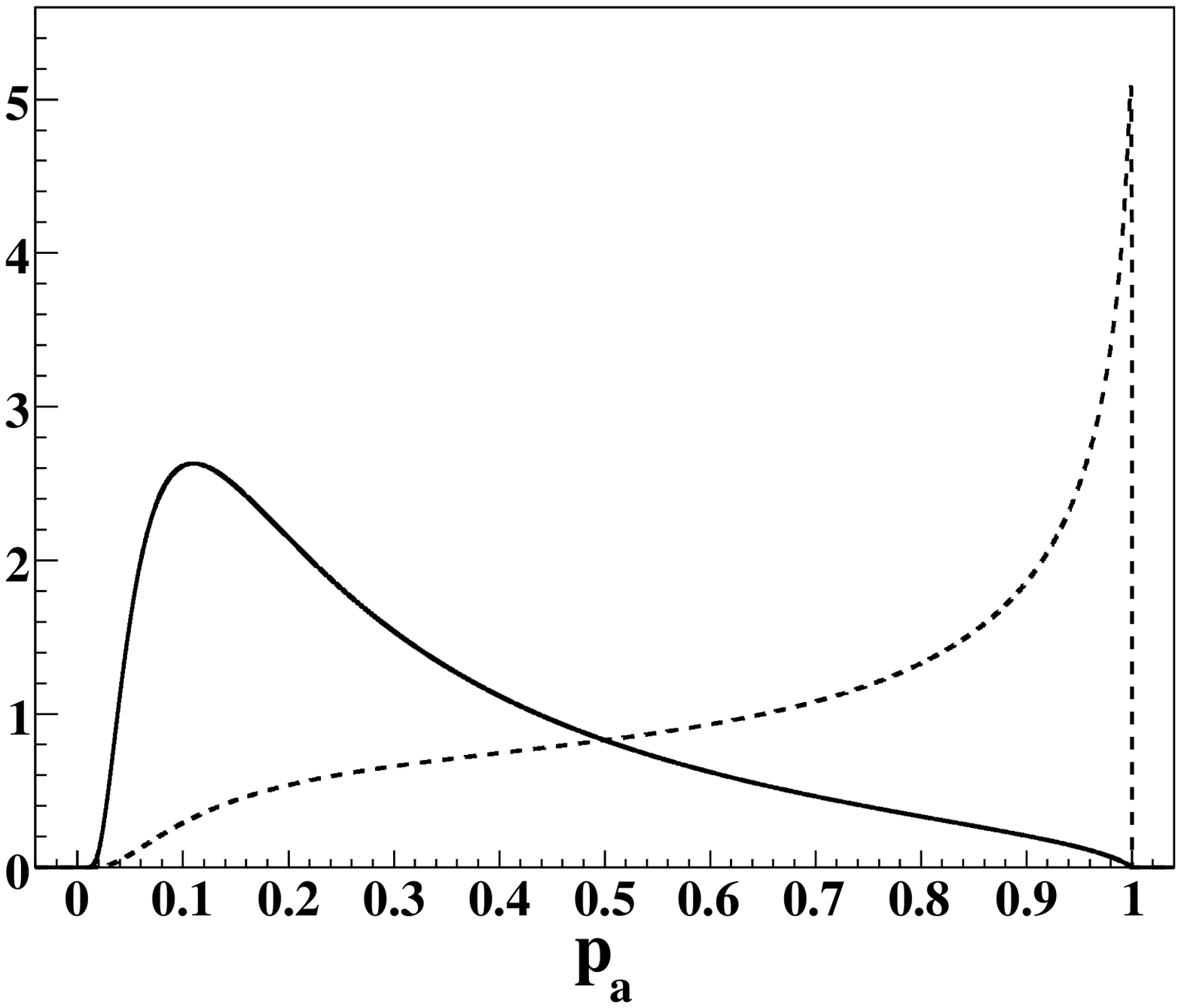}
\includegraphics[width=6.8cm]{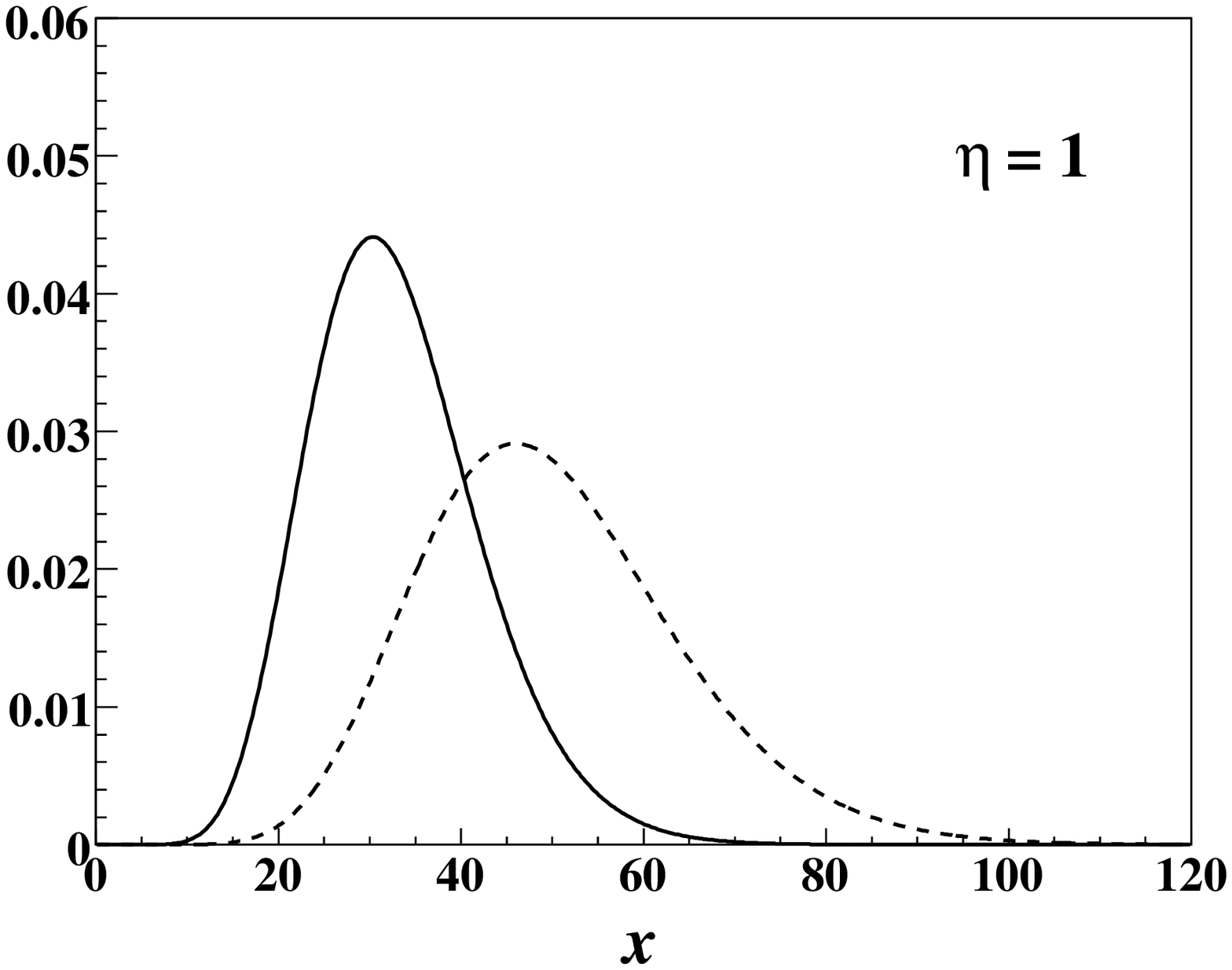}
\includegraphics[width=6.8cm]{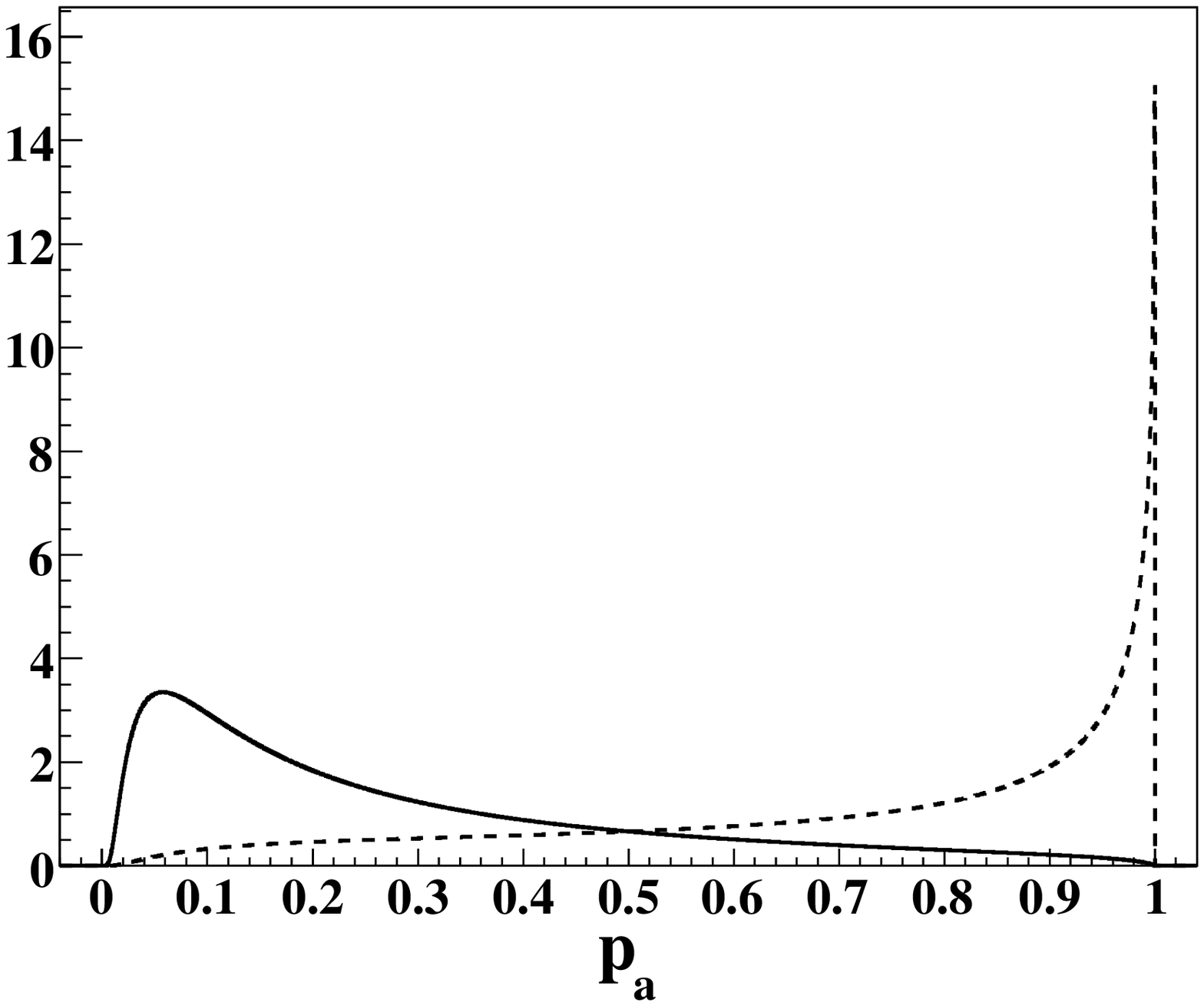}
\includegraphics[width=6.8cm]{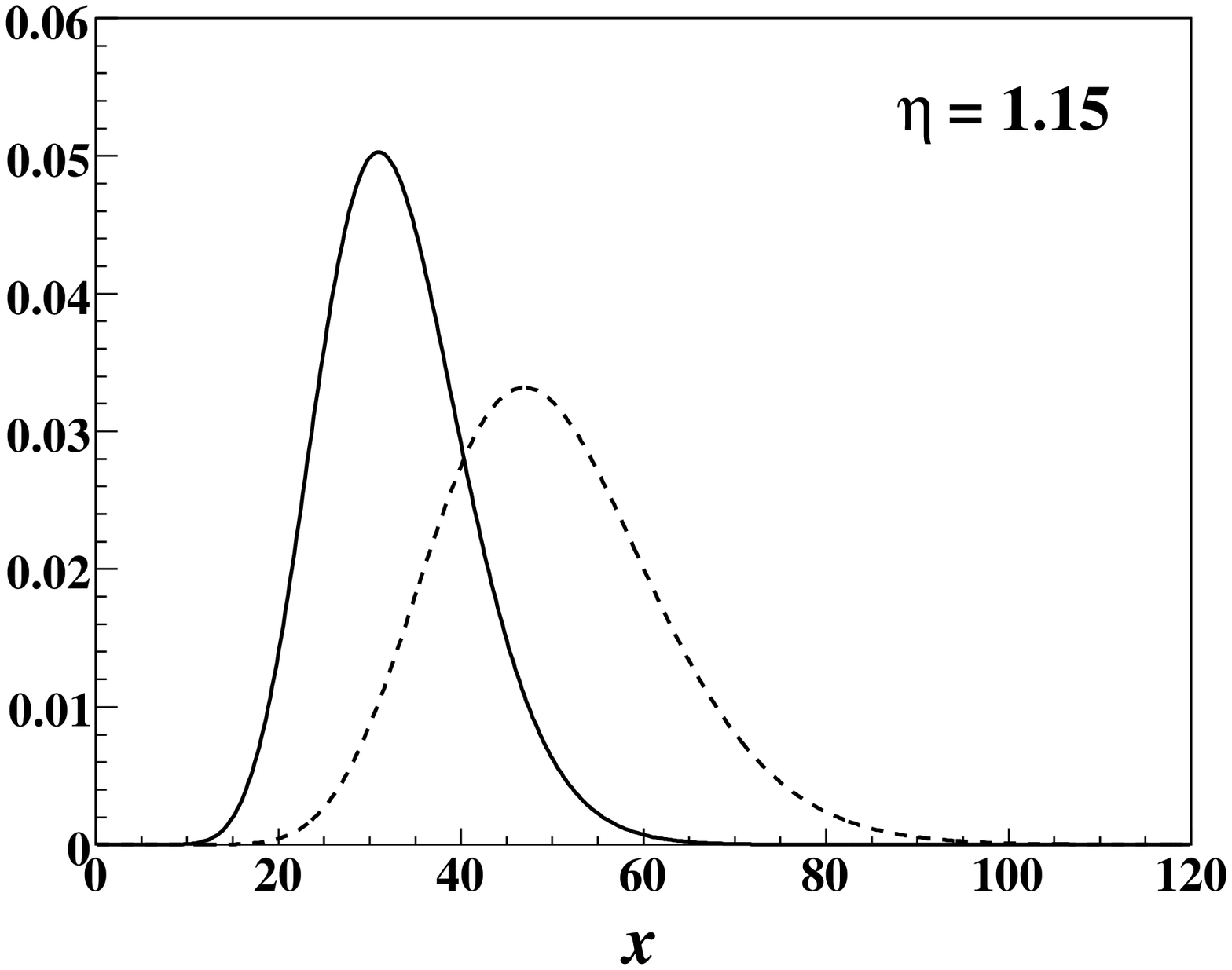}
\includegraphics[width=6.8cm]{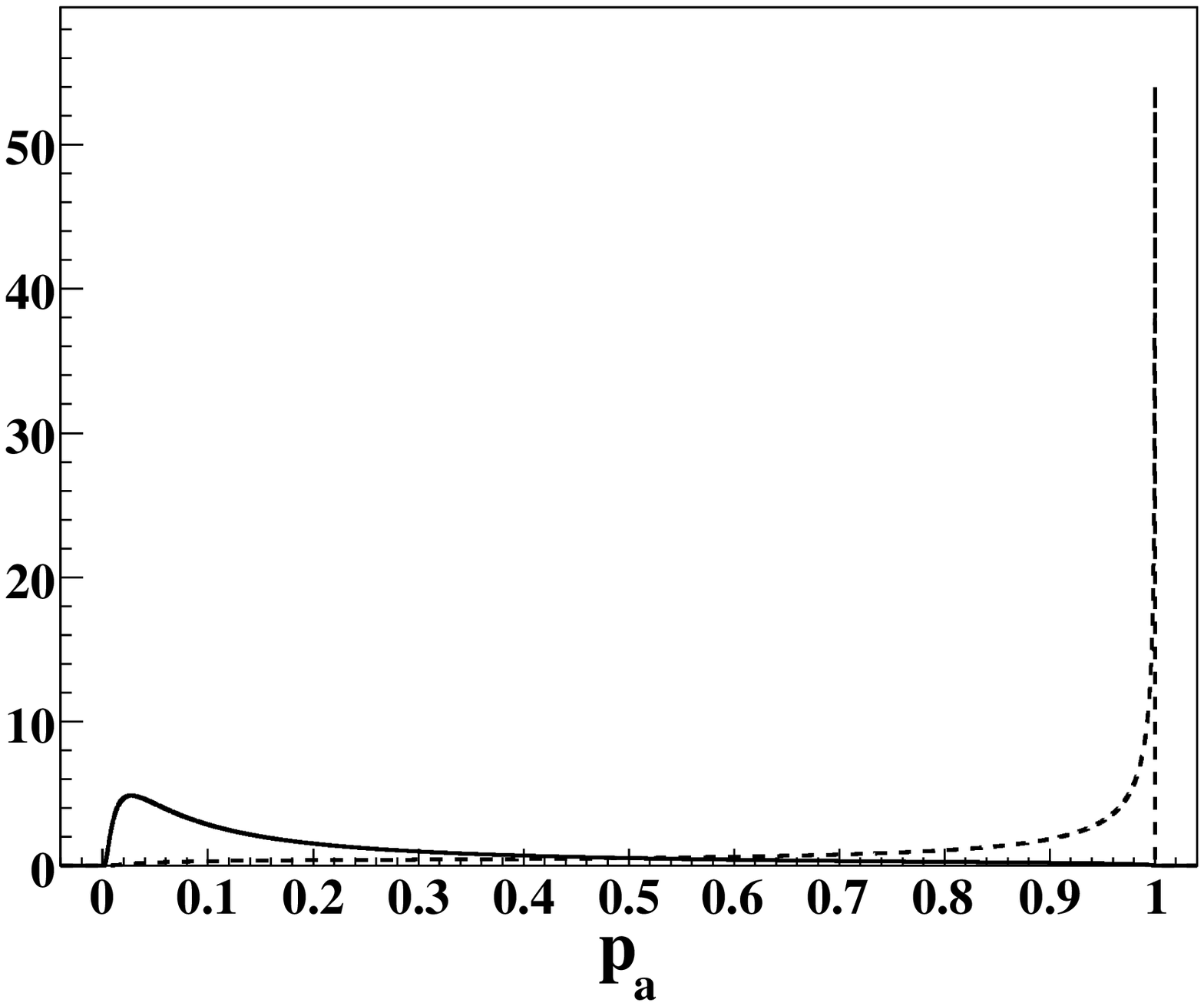}
\includegraphics[width=6.8cm]{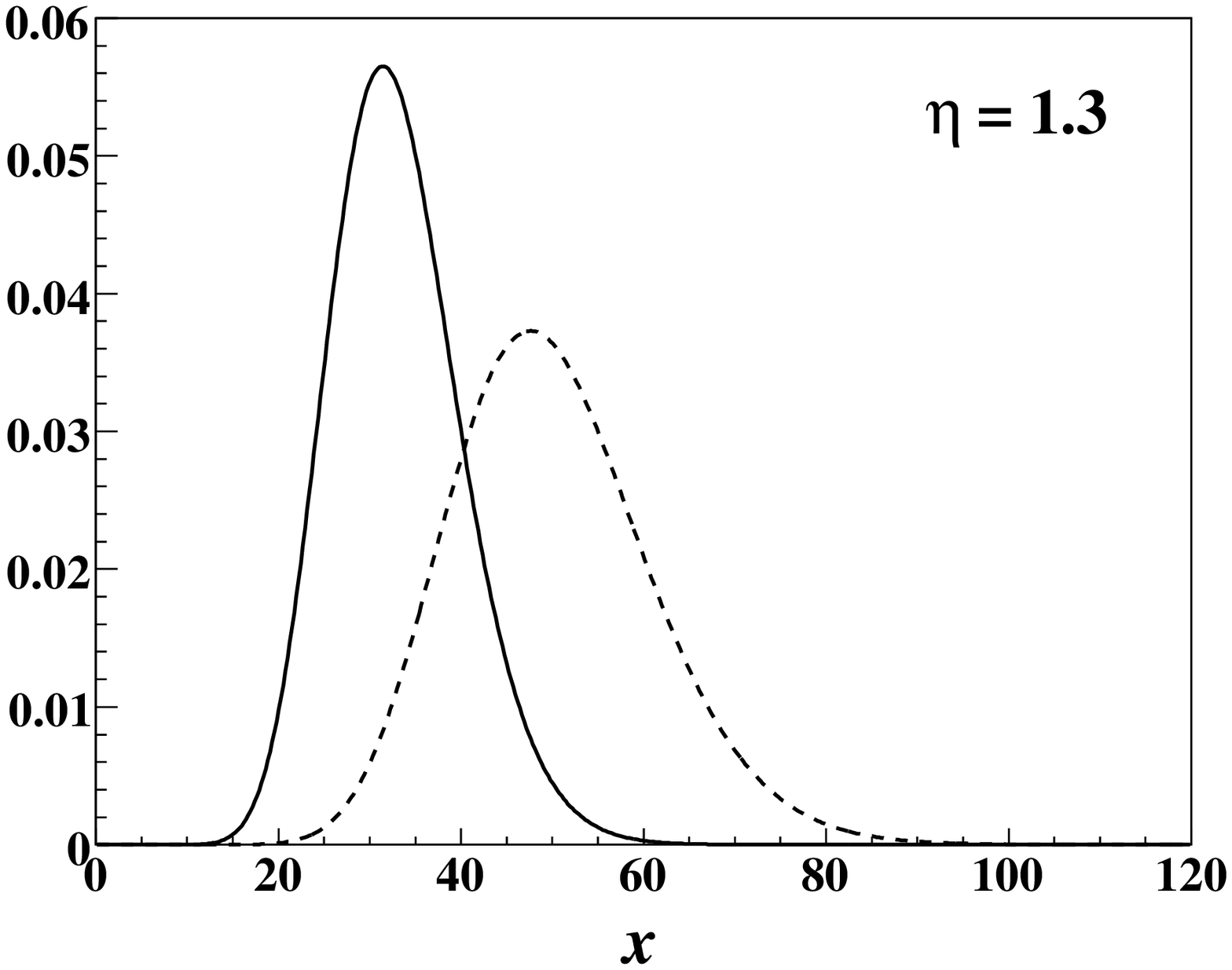}
\caption{Left panels show $P_a(p_a)$ and $P_b(p_a)$ as a function of $p_a$ for three different
values of $\eta$ and right panels show the corresponding distributions functions $f(x;\alpha, \sigma_a)$ 
and $f(x;\alpha, \sigma_b)$. Dashed lines correspond to class $a$ and solid lines to class $b$. The parameters 
$\sigma_a=50$ and $\sigma_b=33$ are fixed and $\alpha$ is varied to obtain the different values of $\eta$ 
considered. \label{Example}}
\end{center}
\end{figure}

Figure \ref{Example} shows $P_a(p_a)$ and $P_b(p_a)$ and the corresponding distributions
$f(x;\alpha, \sigma_a)$ and $f(x;\alpha, \sigma_b)$ for three different values of $\eta$. The 
values of $\eta$ are obtained by fixing $\sigma_a=50$ and $\sigma_b=33$ and changing $\alpha$. 

From figure \ref{Example} it is seen that as $\eta$ increases, the distribution functions $P_a(p_a)$ and $P_b(p_a)$ 
concentrate more around $p_a=1$ and $p_a=0$, respectively, i.e., the classification between classes $a$ and $b$ 
get better. $P_a(p_a)$ and $P_b(p_a)$ are not symmetric, this is due to the different values of the standard 
deviations of $f(x;\alpha, \sigma_a)$ and $f(x;\alpha, \sigma_b)$. Moreover, for the different values of $\eta$ 
considered $P_a(p_a)$ is more concentrated around $p_a=1$ than $P_b(p_a)$ around $p_a=0$, this is because
$\sigma_a > \sigma_b$.  

\begin{figure}[!bt]
\begin{center}
\includegraphics[width=12cm]{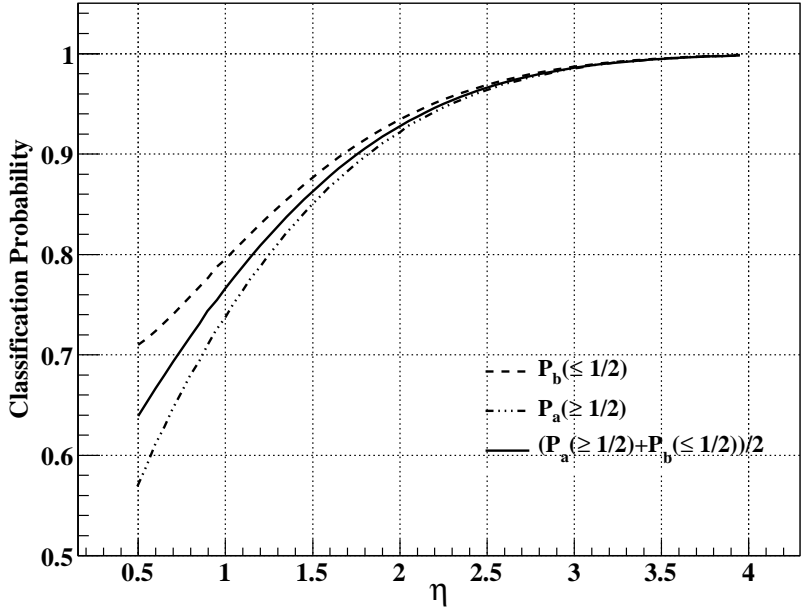}
\caption{Classification probability of classes $a$, $b$ and $a$ and $b$ together as a 
function of $\eta$. \label{ExampleCLP}}
\end{center}
\end{figure}

The classification probabilities for classes $a$ and $b$ are given by,
\begin{eqnarray}
\label{pcla}
P_{cl}^a &=& P_a(p_a \geq 1/2) = \int_{1/2}^1 dp_a\ P_a(p_a), \\
\label{pclb}
P_{cl}^b &=& P_b(p_a \leq 1/2) = \int^{1/2}_0 dp_a\ P_b(p_a), 
\end{eqnarray}
and the classification probability for both classes is $P_{cl}=(P_{cl}^a+P_{cl}^b)/2$. 
Figure \ref{ExampleCLP} shows the classification probability as a function of $\eta$ for 
classes $a$, $b$ and both together. As expected, they increase with $\eta$ getting closer 
to one. 

The classification probability, for the particular distributions considered in this
example, is larger for the distribution with smaller standard deviation, this is not general,
it depends on the shape of the distribution functions considered. To better understand this 
fact the classification probabilities can be rewritten in the following way,
\begin{eqnarray}
P_{cl}^a &=& \int dx\ \Theta(f(x;\alpha, \sigma_a)-f(x;\alpha, \sigma_b)) f(x;\alpha, \sigma_a) \nonumber \\
\label{pclani}
&=& \int_{x_0}^{\infty} dx \ f(x;\alpha, \sigma_a), \\  
P_{cl}^b &=& \int dx\ \Theta(f(x;\alpha, \sigma_b)-f(x;\alpha, \sigma_a)) f(x;\alpha, \sigma_b)  \nonumber \\
\label{pclbni}
&=& \int^{x_0}_{-\infty} dx \ f(x;\alpha, \sigma_b), 
\end{eqnarray}
where $\Theta(x)=1$ for $x \geq 0$ and $\Theta(x)=0$ otherwise and $x_0$ is the solution of 
$f(x_0;\alpha, \sigma_a)=f(x_0;\alpha, \sigma_b)$. The classification probability for a given class is the 
integral of its distribution function over the interval for which it is grater than the distribution function 
of the other class. Therefore, from Eqs. (\ref{pclani}) and (\ref{pclbni}) it can be seen that although it is 
not general that, if the standard deviation of the distribution of class $a$ is grater than the corresponding 
to class $b$ then $P_{cl}^a < P_{cl}^b$, it happens for many different kind of distribution functions, in 
particular, it is true for the ones considered in this work and also for Gaussian distributions.

\section{Simulations}
\label{sim}

\subsection{Optimum energy bin}
\label{OEB}

The primary energy estimated from the experimental data given by an array of Cherenkov detectors is 
obtained by fitting a lateral distribution function to the total signal in each station. This allows 
to interpolate the shower signal at a fixed distance from the core which, in turn, is used as an energy 
estimator. This reference distance is such that the shower fluctuations go through a minimum in its 
vicinity, and its exact value depends on the geometry of the array; for Auger, the reference distance 
is $1000$ m for the $1500$ m baseline spacing and $600$ m for the AMIGA infill of $750$ m of spacing.

The signal at the reference distance is calibrated with the fluorescence telescopes via hybrid events. 
The corresponding energy uncertainty for the 1500 m-array of Auger is $\sim 18\%$ \cite{Roth:07}. Guided 
by this experimental result, we assume in this work a $25\%$ Gaussian energy uncertainty.

The study of the classification probability at energies of order of $E_0 = 10^{18}$ eV requires the 
determination of the energy interval of reconstructed energies, centered at $E_{r0}$, defined as
$\Pi_{r} = [(1-\delta) E_{r0}, (1+\delta) E_{r0}]$ ($\delta = 0.25$ for $25\%$ of energy uncertainty) such
that the fraction of events of the interval $\Pi_{0} = [(1-\delta) E_{0}, (1+\delta) E_{0}]$ is maximum. 
The intervals $\Pi_r$ and $\Pi_0$ are different because of the spectrum, the contamination of events of real 
energies smaller than $10^{18}$ eV which are outside $\Pi_0$ is grater than the one corresponding to energies, 
also outside $\Pi_0$, but above $10^{18}$ eV. We follow the procedure introduced in Ref. \cite{SupaCompo:08} 
to determine $\Pi_{r}$ which is detailed bellow.

The number of showers with real energies between $E$ and $E+dE$ is given by,
\begin{equation}
\frac{dN}{dE}(E) = N_{0}\ (\gamma -1) \frac{E_{1}^{\gamma -1} E_{2}^{\gamma -1}}{%
E_{2}^{\gamma -1}-E_{1}^{\gamma -1}}\ E^{-\gamma},
\label{Spec}
\end{equation}
where $[E_{1}, E_2] = [0.1, 10] \times 10^{18}$ eV is the region of the spectrum considered, $N_0$ is 
the number of events in the interval $[E_1, E_2]$ and $\gamma = 2.7$.

The number of events whose real energies belong to $\Pi_0$, such that the reconstructed energies fall in $\Pi_r$ 
is,
\begin{equation}
f_{A}(E_{r0}) = \int_{(1-\delta) E_{r0}}^{(1+\delta) E_{r0}} dE%
 \int_{(1-\delta) E_{0}}^{(1+\delta) E_{0}} dE' \ \frac{dN}{dE'}(E') \ G(E,E'),
\label{fa}
\end{equation}
where $G(E,E') = \exp[-(E-E')^2/(2 \delta^2 E'^2)]/(\sqrt{2 \pi} \ \delta \ E')$.

On the other hand, the number of events with real energies outside $\Pi_0$ whose reconstructed energies fall
in $\Pi_r$ is,
\begin{eqnarray}
f_{B}(E_{r0}) &=& \int_{(1-\delta) E_{r0}}^{(1+\delta) E_{r0}} dE \left[%
\int_{E_{1}}^{(1-\delta) E_{0}} dE' \ \frac{dN}{dE'}(E') \ G(E,E')+\right. \nonumber \\
&&\left. \int_{(1+\delta) E_{0}}^{E_{2}} dE' \ \frac{dN}{dE'}(E') \ G(E,E')\right].
\label{fb}
\end{eqnarray}

\begin{figure}[!bt]
\begin{center}
\includegraphics[width=12cm]{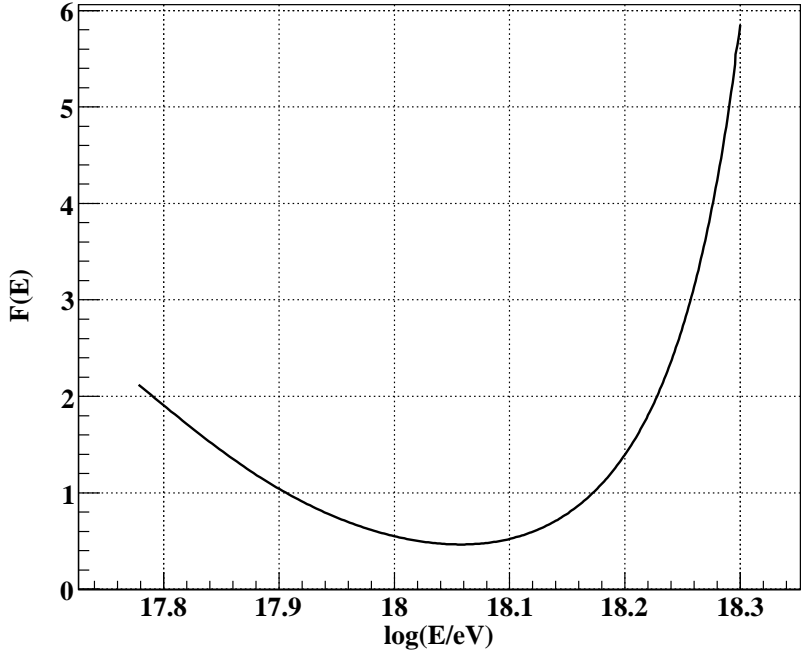}
\caption{Ratio between the number of events whose real energies do not belong to $\Pi_0$ but the reconstructed
one fall in $\Pi_r$ and the number of events whose real energies belong to $\Pi_0$ but the reconstructed one
fall in $\Pi_r$, $F(E_{r0})$. The minimum is located at $E_{r0} \cong 1.14 \times 10^{18}$ eV. \label{BinCenter}}
\end{center}
\end{figure}

Therefore, the value of $E_{r0}$ for which the fraction of events belonging to $\Pi_0$ that fall
in $\Pi_r$ is maximum is obtained by minimizing the function $F(E_{r0})=f_{B}(E_{r0})/f_{A}(E_{r0})$.
Figure \ref{BinCenter} shows $F(E_{r0})$ for $\delta=0.25$ from which can seen that it has a minimum but
at an energy grater than $10^{18}$ eV. The minimum is located at $E_{r0} \cong 1.14 \times 10^{18}$ eV,
then, $\Pi_r =[0.86 ,\ 1.43]\times10^{18}$ eV.

The studies under consideration require the simulation of the cosmic ray energy spectrum in a large energy 
interval around $10^{18}$ eV. The number of showers required to obtain different samples of good statistics is 
very large. This is a very difficult task because of the computer processing time and disk space needed. The 
number of showers is considerable reduced considering just the events whose reconstructed energies fall in 
$\Pi_r$. Therefore, the distribution function of real energies corresponding to events whose reconstructed 
energies fall in $\Pi_r$ can be used to obtain the input energy for the simulations. This distribution is 
given by, 
\begin{equation}
P_{R}(E) = C\ \frac{dN}{dE}(E)\ \int_{(1-\delta) E_{r0}}^{(1+\delta) E_{r0}} dE' \ G(E,E'),
\label{PrDef}
\end{equation}
where,
\begin{equation}
C^{-1} =\int_{E_1}^{E_2} dE \ \frac{dN}{dE}(E)\ \int_{(1-\delta) E_{r0}}^{(1+\delta)% 
E_{r0}} dE' \ G(E,E').
\label{Norm}
\end{equation}
Performing the integral in Eq. (\ref{PrDef}) the following analytical expression is obtained,
\begin{equation}
P_{R}(E) = C\ E^{-\gamma} \left[ \mathrm{Erf}\left( \frac{E+E_{r0} (\delta-1)}{\sqrt{2} \delta E}  \right)-%
\mathrm{Erf}\left( \frac{E-E_{r0} (\delta+1)}{\sqrt{2} \delta E}  \right) \right],
\label{PrRes}
\end{equation}
where,
\begin{equation}
\mathrm{Erf}(z) = \frac{2}{\sqrt{\pi}} \ \int_{0}^{z} dt \ \exp(-t^2).
\label{Erf}
\end{equation}

\begin{figure}[!bt]
\begin{center}
\includegraphics[width=12cm]{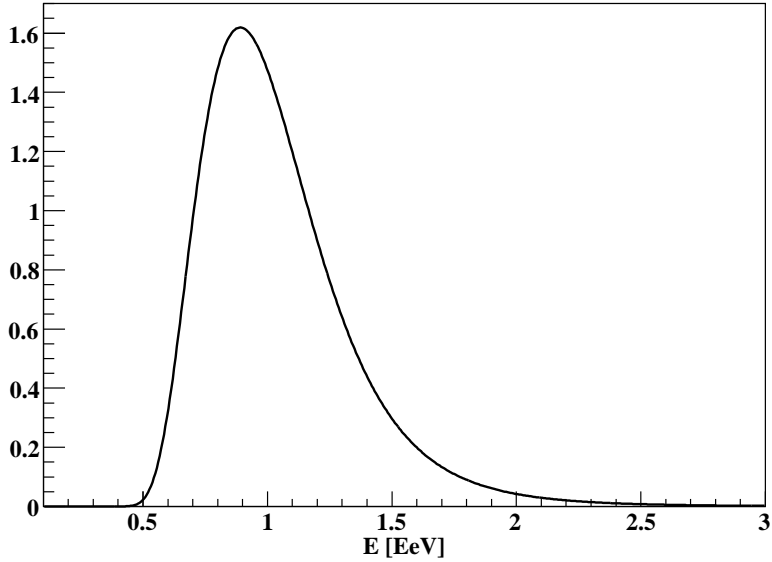}
\caption{Distribution function of real energies for events whose reconstructed energy falls in $\Pi_r$ 
for $\delta=0.25$, $\gamma=2.7$, $E_1=0.1$ EeV and $E_2=10$ EeV. \label{RealDist}}
\end{center}
\end{figure}

Figure \ref{RealDist} shows $P_R(E)$ for $\delta=0.25$, $\gamma=2.7$, $E_1=0.1 \times 10^{18}$ eV and 
$E_2=1\times10^{19}$ eV. The distribution of real energies is not symmetric with respect to $10^{18}$ eV, 
it has a tail at large energies due fundamentally to the assumed constant relative error on primary energy.

\subsection{Showers and detector simulations}
\label{simshdet}

The simulation of the air showers is performed using the package Aires version 2.8.4a \cite{AIRES}.   
Firstly, 8 sets of 20 independent samples are simulated corresponding to proton and iron nuclei as 
primaries, zenith angles $\theta=30^\circ$ and $\theta=40^\circ$ and two different hadronic
interaction models, QGSJET-II \cite{QGIIa,QGIIb} and Sibyll 2.1 \cite{Sib2.1}. Each sample consists 
in 50 showers with primary energies obtained by taking at random 50 independent values from the 
distribution of Eq. (\ref{PrRes}). In this way the effects of the spectrum and energy uncertainty are 
taken into account (see subsection \ref{OEB}). Secondly, 8 samples of 50 showers each are also generated
corresponding to proton and iron primaries, $\theta=20^\circ,\ 25^\circ,\ 35^\circ,\ 45^\circ$ and 
QGSJET-II as the hadronic interaction model. Also in this case the input energy is obtained by sampling 
the distribution of Eq. (\ref{PrRes}).

The simulation of the response of the Cherenkov detectors and the muon counters of the 750 m-AMIGA 
array is performed by using a dedicated package described in Ref. \cite{SupaRec:08}. Muon detectors 
of 30 m$^2$ of area segmented in 192 parts are used for the simulations. It is assumed a time resolution 
of 20 ns and the efficiency of each bar equal to one. Each shower is used 50 times by uniformly distributing 
impact points in the array area. The reconstruction of the arrival direction and core position of the 
events is performed by using the package of Ref. \cite{CDAS}, specially developed to reconstruct the Cherenkov 
detector information in Auger. The muon lateral distribution function is reconstructed using the method 
introduced in Ref. \cite{SupaRec:08} and the parameter $X_{max}$ is obtained following the method described 
in Ref. \cite{SupaRec:08}, which takes into account the effect of the HEAT telescopes and the Reconstruction
procedure of the longitudinal profile.  

In short, 20 independent samples of $N\cong50\times50=2500$ events (depending on the reconstruction efficiency) 
are obtained for each $\theta\in\{30^\circ, 40^\circ\}$, type of primary and hadronic interaction model. Besides, 
two samples of $N\cong2500$ events are also obtained for each $\theta\in\{20^\circ,\ 25^\circ,\ 35^\circ,\ 45^\circ\}$ 
type of primary and QGSJET-II as the hadronic interaction model.

\section{Numerical approach}
\label{clprop}

\subsection{Methodology and results}
\label{MethRes}

The classification method considered in this work require the knowledge of the conditional density
functions $P(\vec{x}\ | A)$. Therefore, the non-parametric method of kernel superposition 
\cite{Silvermann:86,Scott:92,Fadda:98,Merritt:94} is used to estimate the probability density 
functions, from the simulated data, for the different set of mass sensitive parameters considered. 
Besides, the adaptive bandwidth method introduced by B. Silverman \cite{Silvermann:86} is also 
implemented in order to obtain better estimates of the density functions. 

The procedure starts by performing a first estimation of each density function using a Gaussian 
kernel with fixed smoothing parameter,
\begin{equation}
\hat{P}_{0}(\vec{x}\ | A) = \frac{1}{N \sqrt{|\mathbf{V}|}\ (2 \pi)^{d/2}\ h_{0}^d} \sum_{i=1}^{N}\ %
\exp\left[ -\frac{(\vec{x}-\vec{x}_i)^T \mathbf{V}^{-1} (\vec{x}-\vec{x}_i)}{2 h^2_{0}} \right],
\label{fest0}
\end{equation}
where $\vec{x}$ is a $d$-dimensional vector corresponding to one of the sets of parameters sensitive to 
the primary mass considered, $N$ is the size of the sample, $\mathbf{V}$ is the covariance matrix of 
the data sample and $h_{0} = 1.06 \times N^{-1/(d+4)}$ is the smoothing parameter corresponding to Gaussian 
samples which is used very often in the literature because it gives very good estimates even for non 
Gaussian samples. 

The following parameters are calculated by using the estimate obtained from Eq. (\ref{fest0}),
\begin{equation}
\lambda_{i} = \left[ \frac{\hat{P}_{0}(\vec{x}_{i}\ |A)}{\left( \prod_{j=1}^{N} \hat{P}_{0}(\vec{x}_{j}\ |A)%
\right)^{1/N} } \right]^{-1/2},
\label{lamda}
\end{equation}
and then, the final density estimate is obtained from,
\begin{equation}
\hat{P}(\vec{x}\ |A) = \frac{1}{N \sqrt{|\mathbf{V}|}\ (2 \pi)^{d/2}} \sum_{i=1}^{N}\ \frac{1}{h^d_{i}}%
\exp\left[ -\frac{(\vec{x}-\vec{x}_i)^T \mathbf{V}^{-1} (\vec{x}-\vec{x}_i)}{2 h^2_{i}} \right],
\label{fest}
\end{equation}
where $h_{i} = h_{0} \ \lambda_{i}$.

As mentioned, different type of mass sensitive parameters are used for the subsequent analyses: the depth of 
the maximum development of the shower, $X_{max}$, the number of muons at 600 m from the shower axis, 
$N_\mu(600)$ and a set of parameters coming from the Cherenkov detectors, $SD=\{t_{1/2},\ \beta,\ R \}$, 
where $t_{1/2}$ is a parameter constructed from the individual rise-time of a subset of stations belonging 
to each event (see Ref. \cite{SupaRec:08}), $\beta$ is the slope of the lateral distribution function of 
the signal in the Cherenkov detectors and $R$ is the curvature radius of the shower front. The following 
combination of parameters are considered: ($i$) $(N_\mu(600),X_{max})$, ($ii$) $SD=(t_{1/2},\beta,R)$, 
($iii$) $(N_\mu(600),SD)$, ($iv$) $(X_{max},SD)$ and ($v$) $All=(N_\mu(600),X_{max},SD)$.

Ten pairs of density estimates, $\{\hat{P}_i(\vec{x}\ |Pr),\hat{P}_i(\vec{x}\ |Fe)\}$ with $i=1\ldots 10$
are obtained for each zenith angle ($\theta=30^\circ$ and $40^\circ$), hadronic interaction model and 
set of parameters. 10 of the 20 samples (see subsection \ref{sim}) corresponding to protons and 10 of the 20 
samples corresponding to iron nuclei are used to construct those estimates. The rest 10 proton samples and 10 
iron samples for each zenith angle and hadronic interaction model are left as test samples.

Therefore, 10 different estimates of the probability of proton given $\vec{x}$ are obtained 
(see Eq. (\ref{Bayes})),
\begin{equation}
\hat{P}_i(Pr\ |\vec{x})=\frac{\hat{P}_i(\vec{x}\ |Pr)}{\hat{P}_i(\vec{x}\ |Pr)+\hat{P}_i(\vec{x}\ |Fe)}%
\ \ \textrm{with} \ \ i=1 \ldots 10,
\label{pprx}
\end{equation}
assuming no prior knowledge.

Each $\hat{P}_i(Pr\ |\vec{x})$ in combination with the 20 proton and iron test samples are used
to obtain samples of the distributions functions, $P_{pr}(p_{pr})$ and $P_{fe}(p_{pr})$, where 
$p_{pr}$ is the random variable obtained evaluating $P_i(Pr\ |\vec{x})$ with a random vector 
$\vec{x}$ distributed following the proton or iron distributions (see Eq. (\ref{pa})). In this way,
100 samples of the distributions $P_{pr}(p_{pr})$ and $P_{fe}(p_{pr})$ are obtained for each zenith 
angle, hadronic interaction model and set of parameters considered.

\begin{figure}[!bt]
\begin{center}
\includegraphics[width=6.8cm]{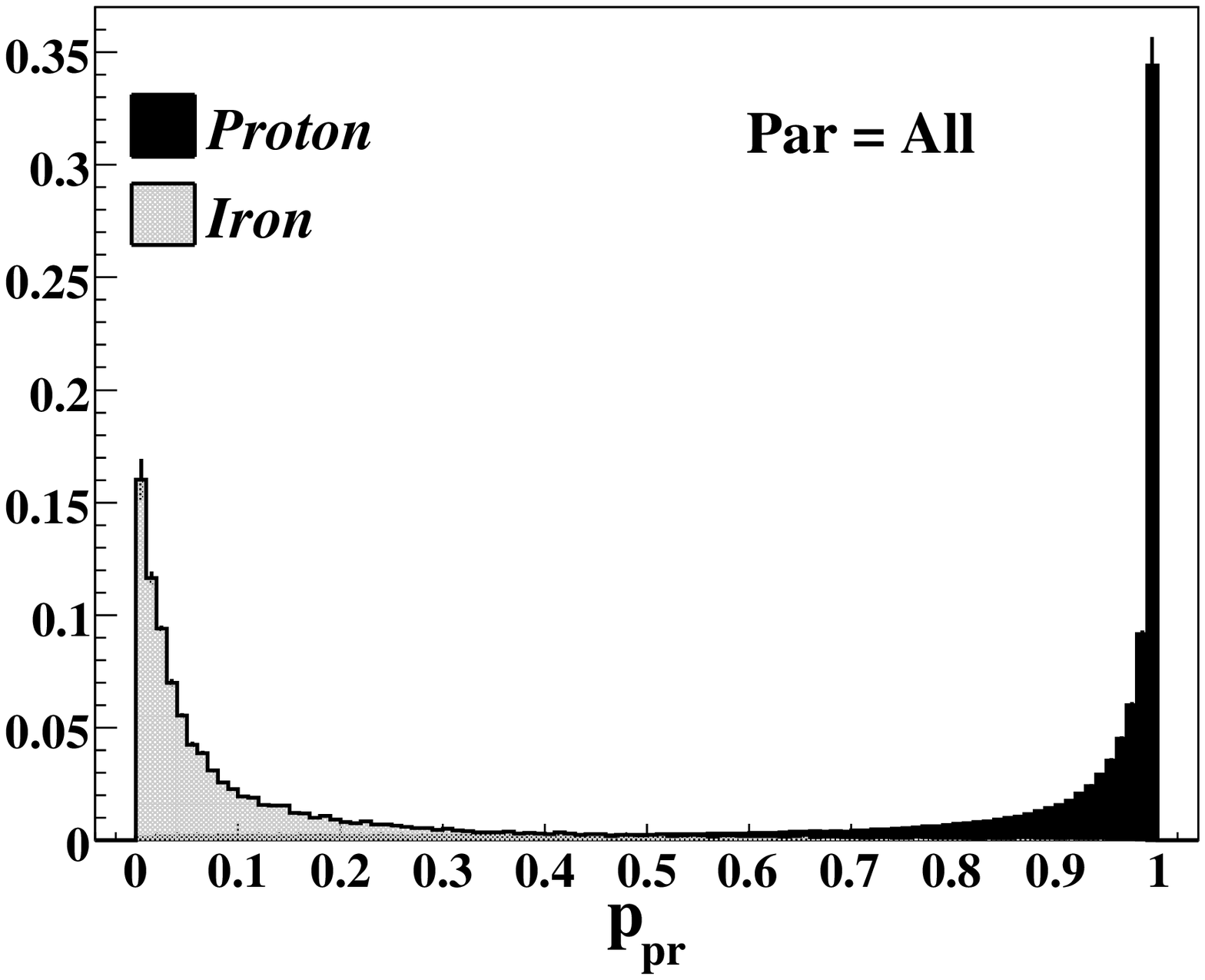}
\includegraphics[width=6.8cm]{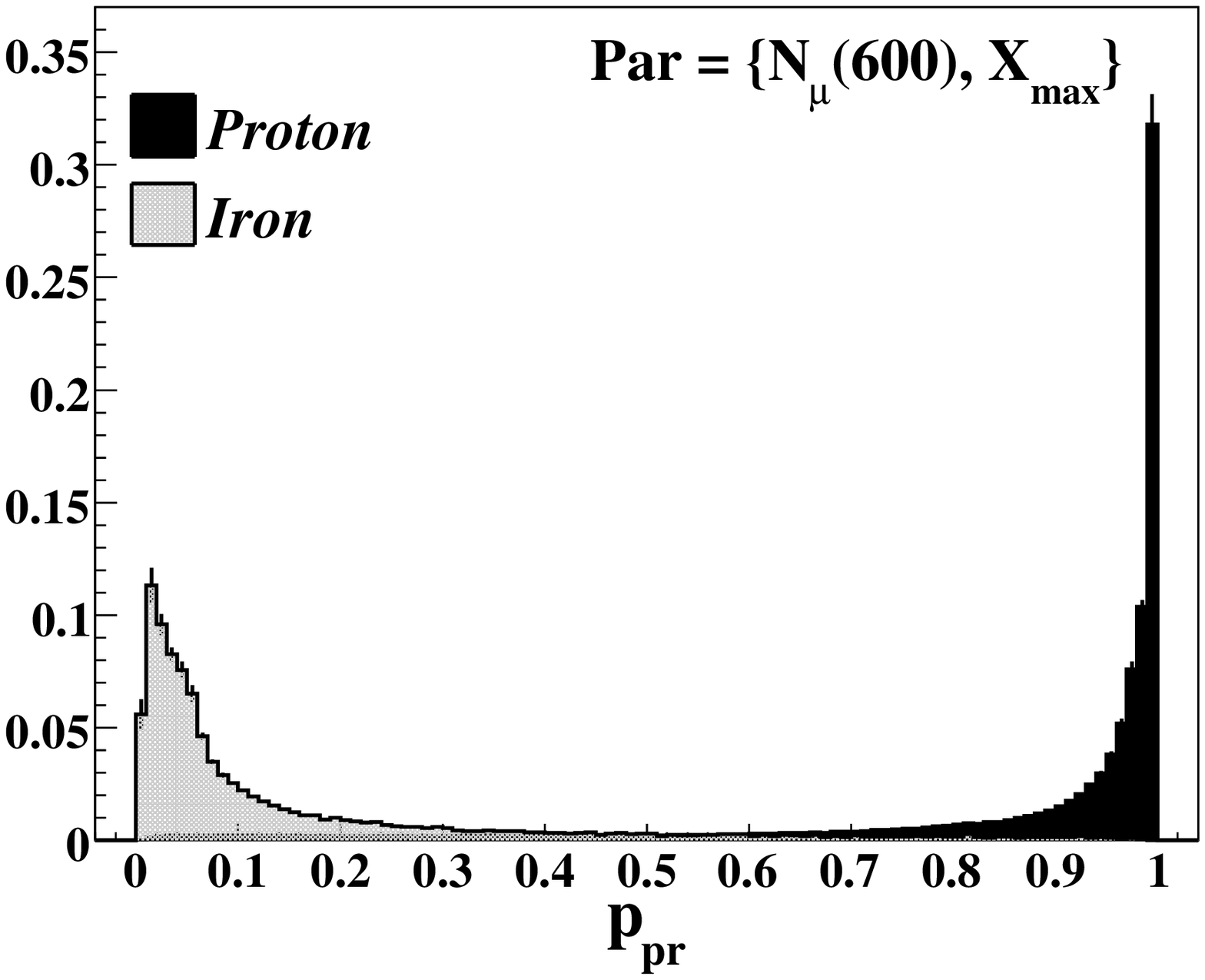}
\includegraphics[width=6.8cm]{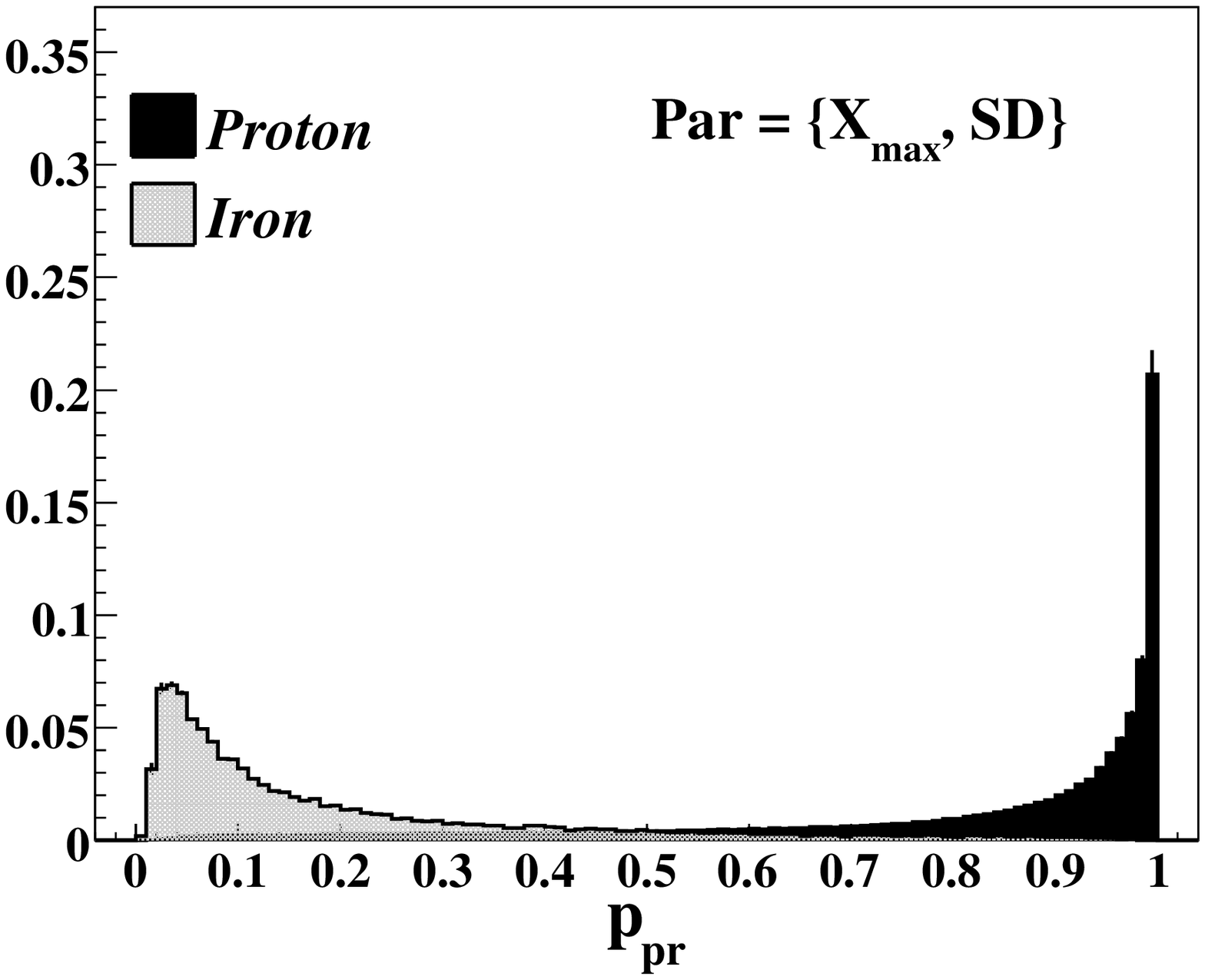}
\includegraphics[width=6.8cm]{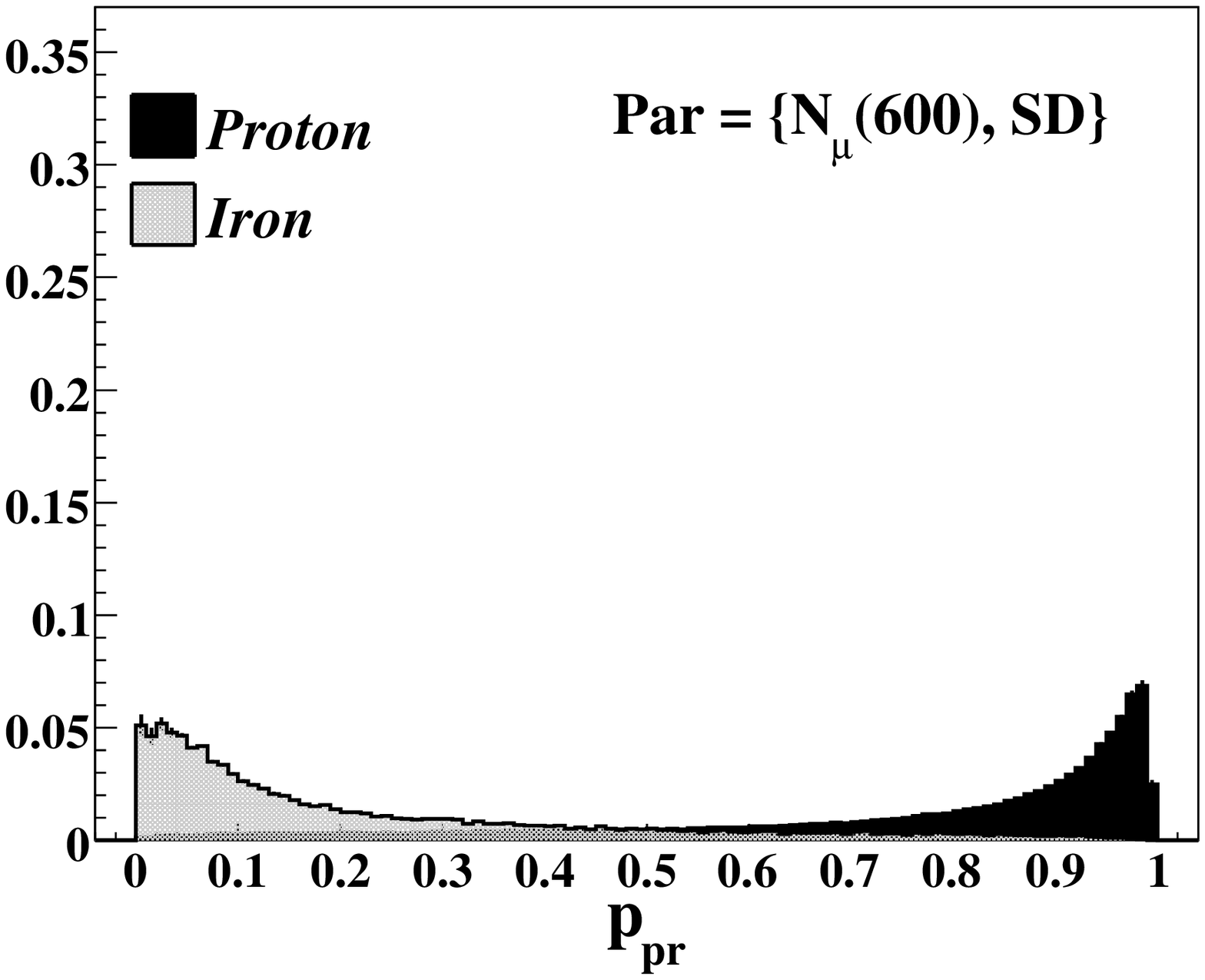}
\includegraphics[width=6.8cm]{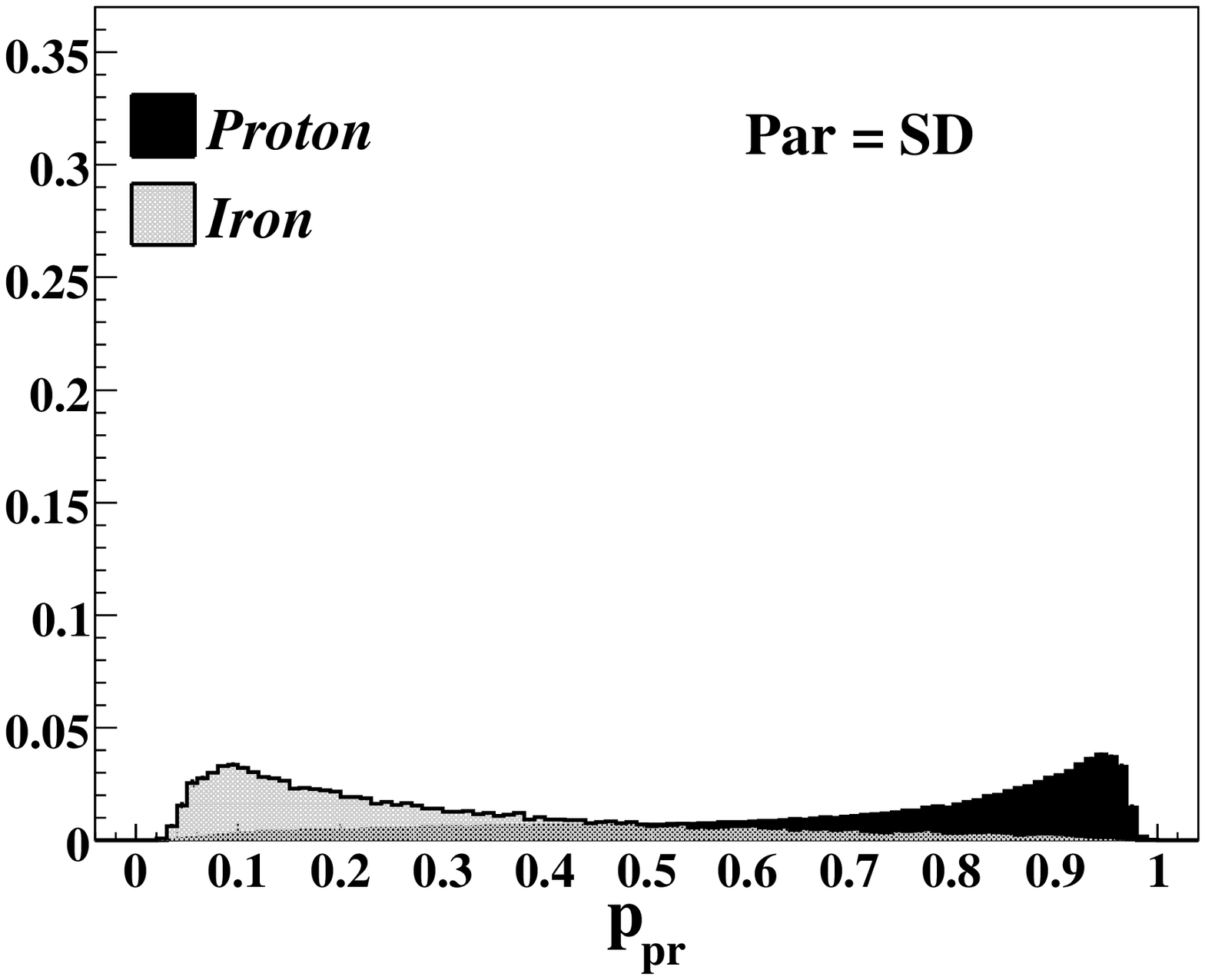}
\caption{Average distributions of $p_{pr}$ with the corresponding errors for protons and iron nuclei and
for the different sets of parameters considered. The zenith angle is $\theta=30^\circ$ and the 
hadronic interaction model QGSJET-II. \label{PprDist}}
\end{center}
\end{figure}

Figure \ref{PprDist} shows the average, with the corresponding errors, of the 100 histograms 
of the samples of the distributions $P_{pr}(p_{pr})$ and $P_{fe}(p_{pr})$ corresponding to 
$\theta = 30^\circ$, QGSJET-II and for the different combinations of parameters considered. It shows
that the best event by event classification is obtained by using all parameters because the distribution 
for protons is the most concentrated around one and the corresponding to iron nuclei around zero. The 
following best combination of parameters is $(N_\mu(600),X_{max})$ but not too far from the previous case. 
Although the combination $(X_{max},SD)$ is better than $(N_\mu(600),SD)$, the improvement in the classification 
probability given by the addition of $N_\mu(600)$ to the other surface detector parameters is very 
important because the duty cycle of the fluorescence detectors is $\sim 10\%$, which means that the size 
of the data sample with available $X_{max}$ is $\sim 10\%$ of the corresponding to the surface detectors. 
Moreover, it is assumed, as an upper bound, $25\%$ of energy uncertainty, presumably the energy uncertainty 
will be smaller, like in the higher energy range which is $\lesssim 20\%$. In this case the classification 
probability of the combination $(N_\mu(600),SD)$ will be equal or even better (depending on primary energy) 
than $(X_{max},SD)$ (see Ref. \cite{SupaRec:08}).             

Figure \ref{PprDist} also shows that the distribution of $p_{pr}$ for protons are more 
concentrated around one than the corresponding to iron nuclei around zero. This is due to, 
for most of the parameters considered\footnote{This is not the case for $N_\mu(600)$, for which,
although the shower to shower fluctuations in the muon content are smaller for iron nuclei, when 
the energy uncertainty is included, they become of the same order or even larger than the ones for 
protons.}, the fluctuations corresponding to protons are larger, like in the simplified example 
of section \ref{clastech}.  

The classification probability for protons, iron nuclei and both together are estimated from
(see Eqs. (\ref{pcla},\ref{pclb})),
\begin{eqnarray}
\label{pclpr}
P_{cl}^{pr}&=&\frac{1}{N_{pr}}\ \sum_{i=1}^{N_{pr}} \Theta(p_{pr,i}^{pr}-1/2) = \frac{n_{pr}}{N_{pr}},  \\
\label{pclfe}
P_{cl}^{fe}&=&\frac{1}{N_{fe}}\ \sum_{i=1}^{N_{fe}} \Theta(1/2-p_{pr,i}^{fe}) = \frac{n_{fe}}{N_{fe}},  \\
\label{pcl}
P_{cl}&=&\frac{N_{pr}\ P_{cl}^{pr} + N_{fe}\ P_{cl}^{fe}}{N_{pr}+N_{fe}},
\end{eqnarray}
where $p_{pr,i}^{A}$ is the probability of $i$th event, corresponding to a sample        
of primary type $A=pr,fe$ and size $N_{A}$, to belong to the proton class and $n_{A}$
is the number of events correctly classified.  

\begin{figure}[!bt]
\begin{center}
\includegraphics[width=6.8cm]{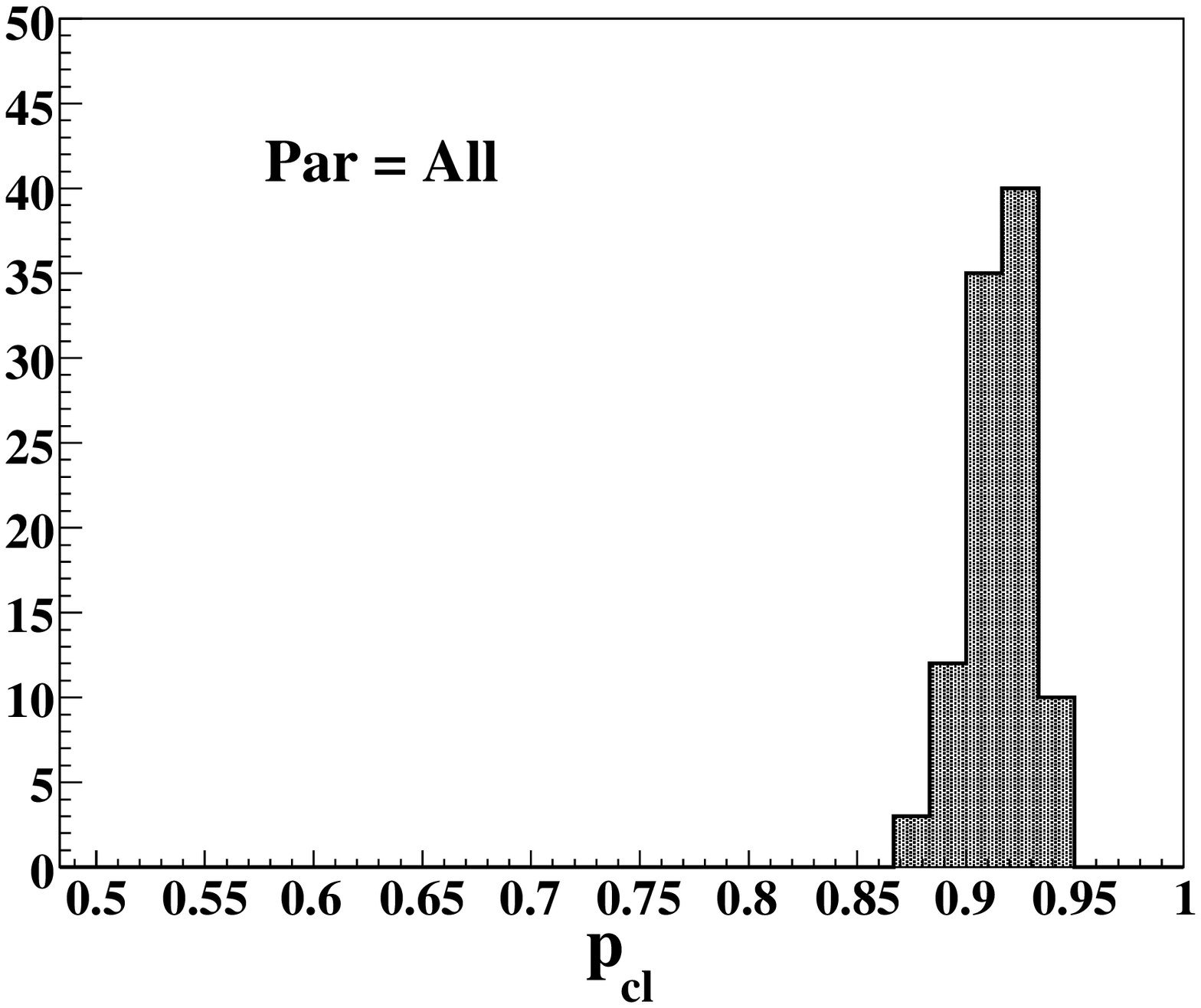}
\includegraphics[width=6.8cm]{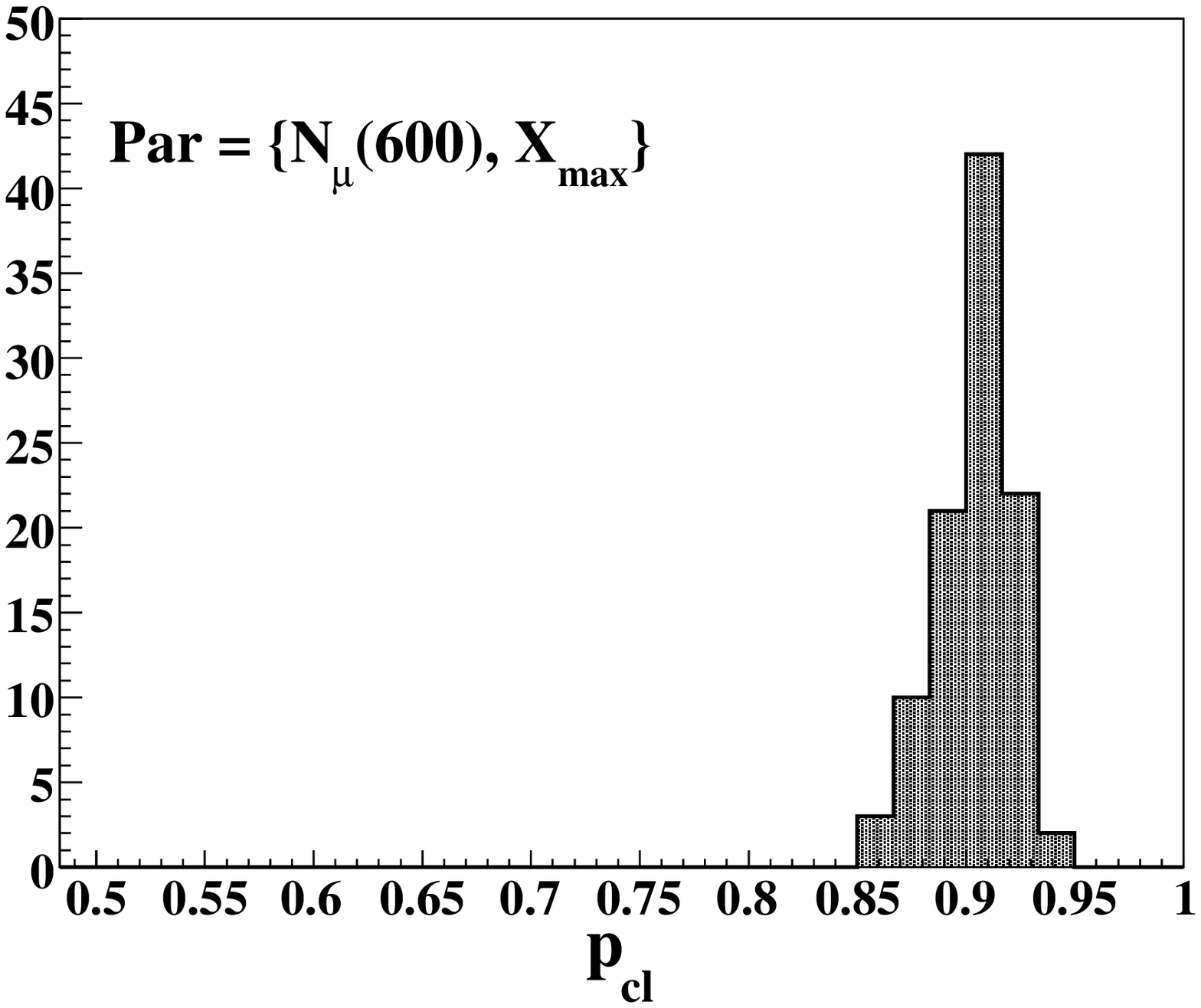}
\includegraphics[width=6.8cm]{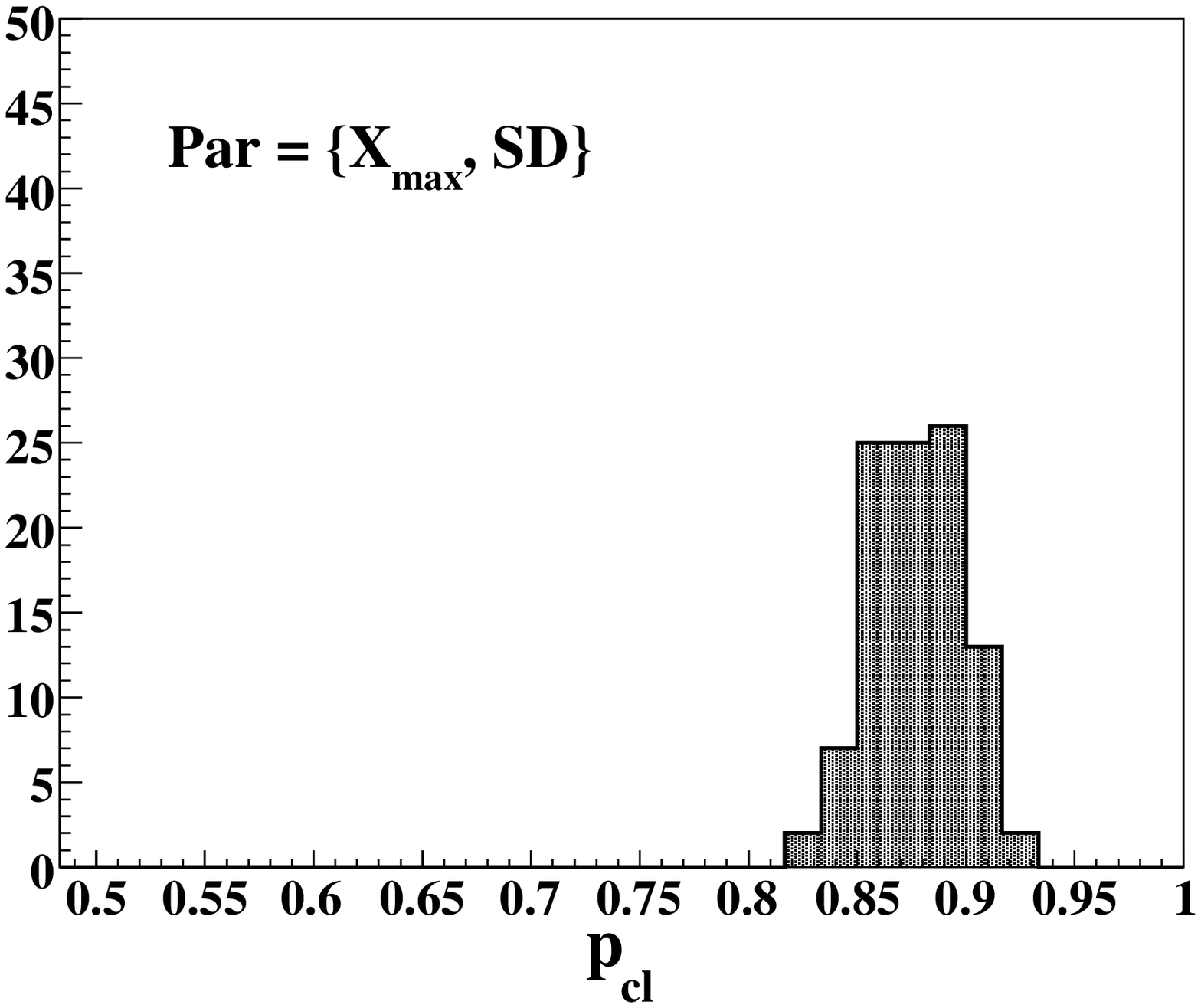}
\includegraphics[width=6.8cm]{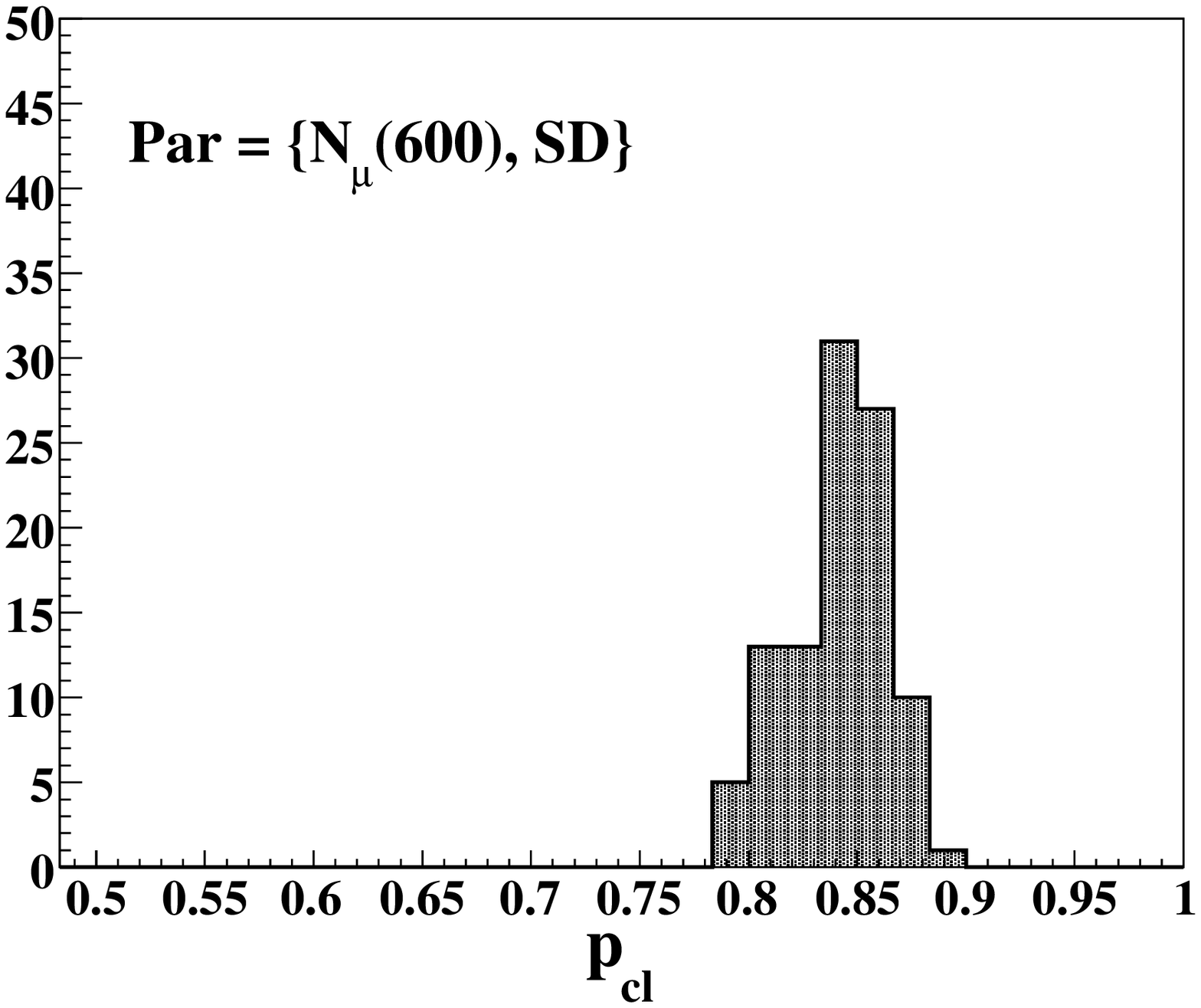}
\includegraphics[width=6.8cm]{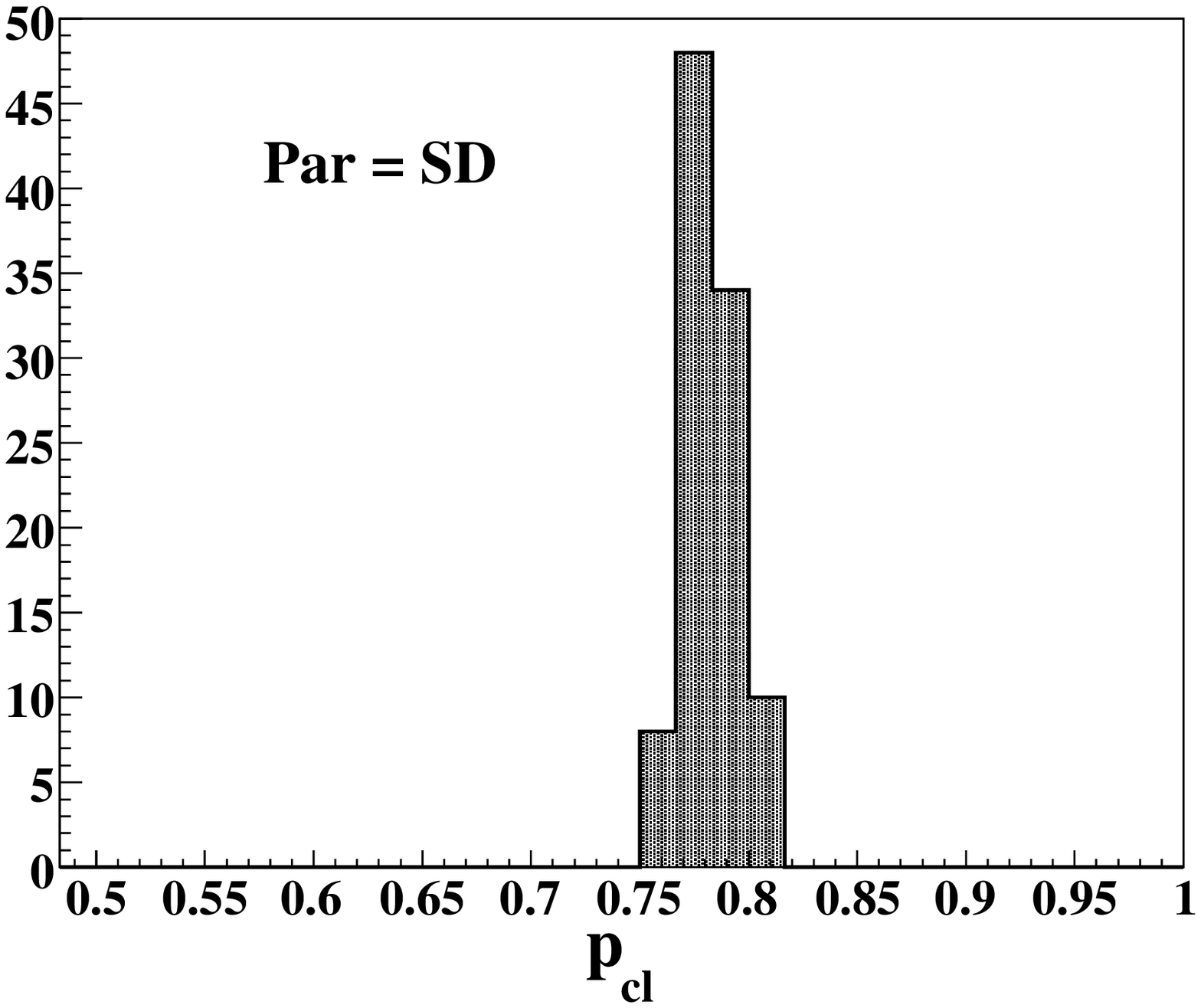}
\caption{Distributions of $P_{cl}$ for $\theta=30^\circ$ and QGSJET-II as the hadronic interaction model
for the different sets of parameters considered.
\label{PclDist}}
\end{center}
\end{figure}

Figure \ref{PclDist} shows the distributions of $P_{cl}$ for QGSJET-II, $\theta=30^\circ$ and each set
of parameters considered. Consistently with the results obtained for distribution functions of $p_{pr}$, 
the highest classification probability is obtained using all parameters considered followed in 
decreasing order by $(N_\mu(600),X_{max})$, $(X_{max},SD)$, $(N_\mu(600),SD)$ and finally $SD$.

Table \ref{tpclqq30} shows the medians and the regions of $68\%$ of probability for the distributions of
figure \ref{PclDist} and also for the corresponding to proton and iron classification probability. Although the 
distributions of $p_{pr}$ for protons are more concentrated around one than the corresponding to iron around zero, 
the classification probabilities for iron nuclei are in general grater than for protons. This happens because, the 
fluctuations of the proton parameters are in general larger than for iron nuclei (see example of section 
\ref{clastech}).    
\begin{table}[h]
\begin{center}
\caption{Medians and regions of $68\%$ of probability for the classification probability distribution
of protons, iron nuclei and both together, for $\theta = 30^\circ$ and hadronic interaction model
QGSJET-II.}
\label{tpclqq30}
\begin{tabular}{c c c c} \hline
Parameters      & $P_{cl}^{pr}$ &  $P_{cl}^{fe}$ & $P_{cl}$       \\   \hline 
$All$           & $0.91^{+ 0.02}_{-0.03}$ &  $0.92^{+0.03}_{-0.01}$ & $0.92^{+0.01}_{-0.02}$ \\ 
$N_\mu+X_{max}$ & $0.90^{+ 0.03}_{-0.04}$ &  $0.90^{+0.04}_{-0.01}$ & $0.91\pm0.02$          \\  
$X_{max}+SD$    & $0.86^{+ 0.04}_{-0.03}$ & $0.89 \pm 0.02$         & $0.88 \pm 0.02$        \\  
$N_\mu+SD$      & $0.84^{+0.03}_{-0.07}$  &  $0.86^{+0.02}_{-0.03}$ & $0.85^{+0.03}_{-0.03}$ \\  
$SD$            & $0.77 \pm0.03$          &  $0.80^{+0.01}_{-0.03}$ & $0.78 \pm 0.01$\\  \hline
\end{tabular}
\end{center}
\end{table}

\begin{figure}[!bt]
\begin{center}
\includegraphics[width=6.8cm]{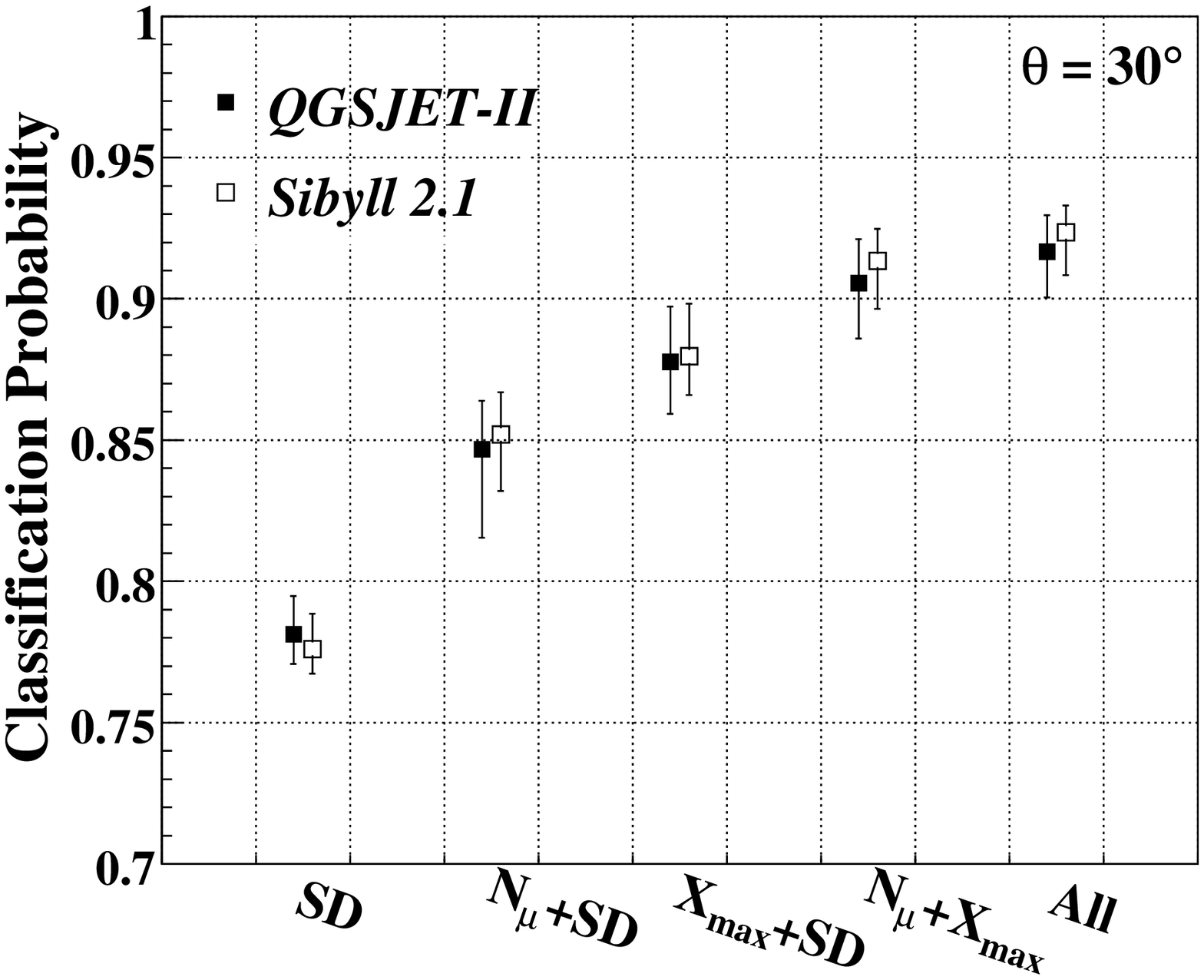}
\includegraphics[width=6.8cm]{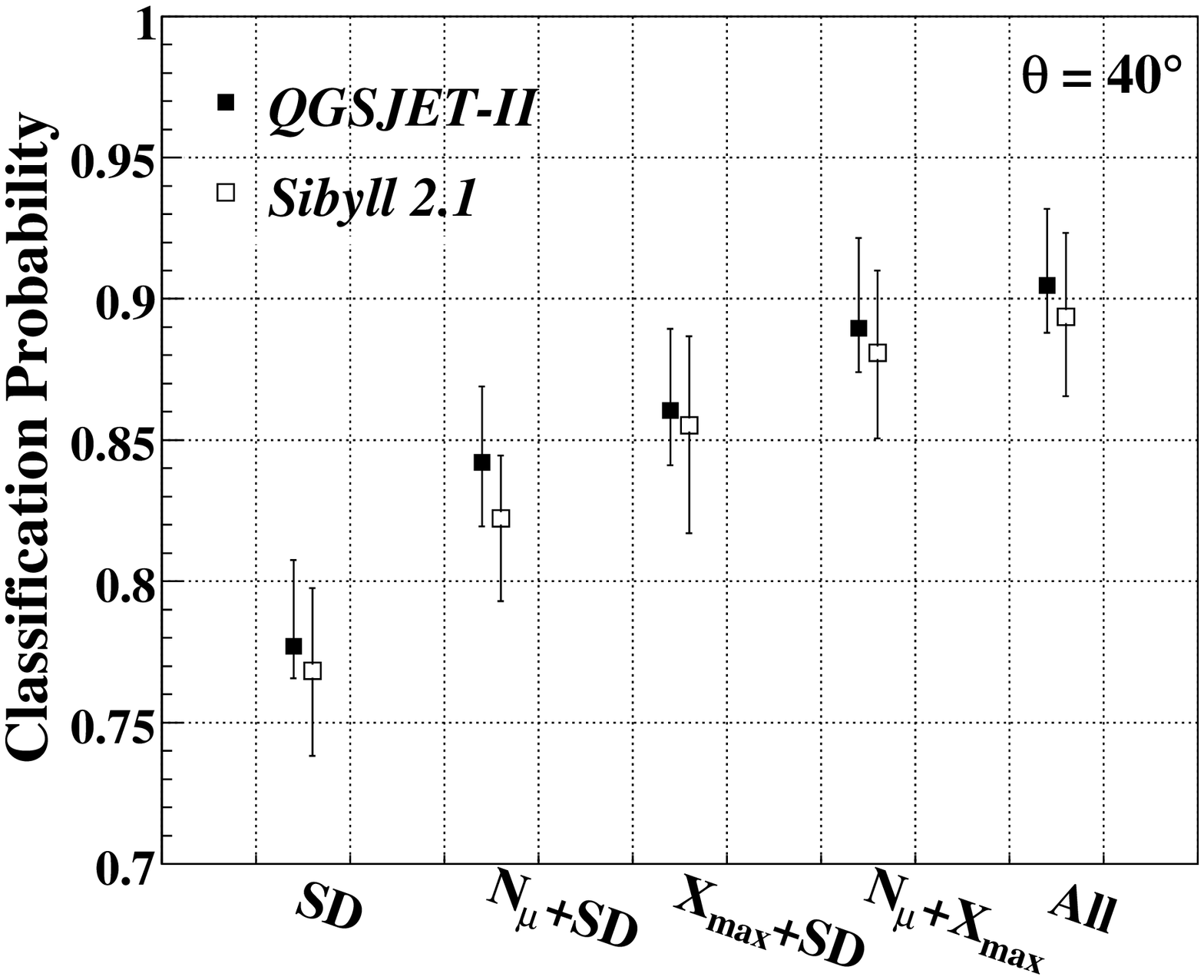}
\caption{Classification probability (medians and regions of $68\%$ of probability) for the different 
sets of parameters considered, for $\theta = 30^\circ$ and $\theta = 40^\circ$ and for the hardronic 
interaction models QGSJET-II and Sibyll 2.1.\label{PclMet1}}
\end{center}
\end{figure}

Figure \ref{PclMet1} shows the medians and the region of $68\%$ of probability for the  
classification probability distributions corresponding to the different combinations of 
parameters, $\theta = 30^\circ$ and $\theta = 40^\circ$ and for the hardronic interaction 
models QGSJET-II and Sibyll 2.1. Although the differences between the distributions corresponding 
to QGSJET-II and Sibyll 2.1 are small, the medians corresponding to $\theta = 40^\circ$ and 
QGSJET-II are systematically larger than the corresponding to Sibyll 2.1. Besides, the 
classification probabilities are in general larger for $\theta = 30^\circ$, this is due to the 
fluctuation of the parameters are larger for $\theta = 40^\circ$.  

The effect of assuming a given hadronic interaction model as the true one whereas the real one is the other
is also studied. For that purpose, 20 pairs $\{\hat{P}_i(\vec{x}\ |Pr),\hat{P}_i(\vec{x}\ |Fe)\}$ with 
$i=1\ldots 20$ are constructed with the proton and iron samples corresponding to QGSJET-II for $\theta = 30^\circ$ 
and $\theta = 40^\circ$. Then, the corresponding 20 proton samples and 20 iron samples, but generated with 
Sibyll 2.1, are used as test samples. In this way, the relevant distributions are obtained but assuming that 
QGSJET-II is the true hadronic interaction model whereas the real one is Sibyll 2.1 (case $Q-S$). The 
same strategy is repeated but in this case assuming that Sibyll 2.1 is the true hadronic interaction model 
whereas the real one is QGSJET-II (case $S-Q$), i.e., the density estimates are obtained using Sibyll 2.1 samples 
and QGSJET-II samples are used as test samples. 

\begin{figure}[!bt]
\begin{center}
\includegraphics[width=6.8cm]{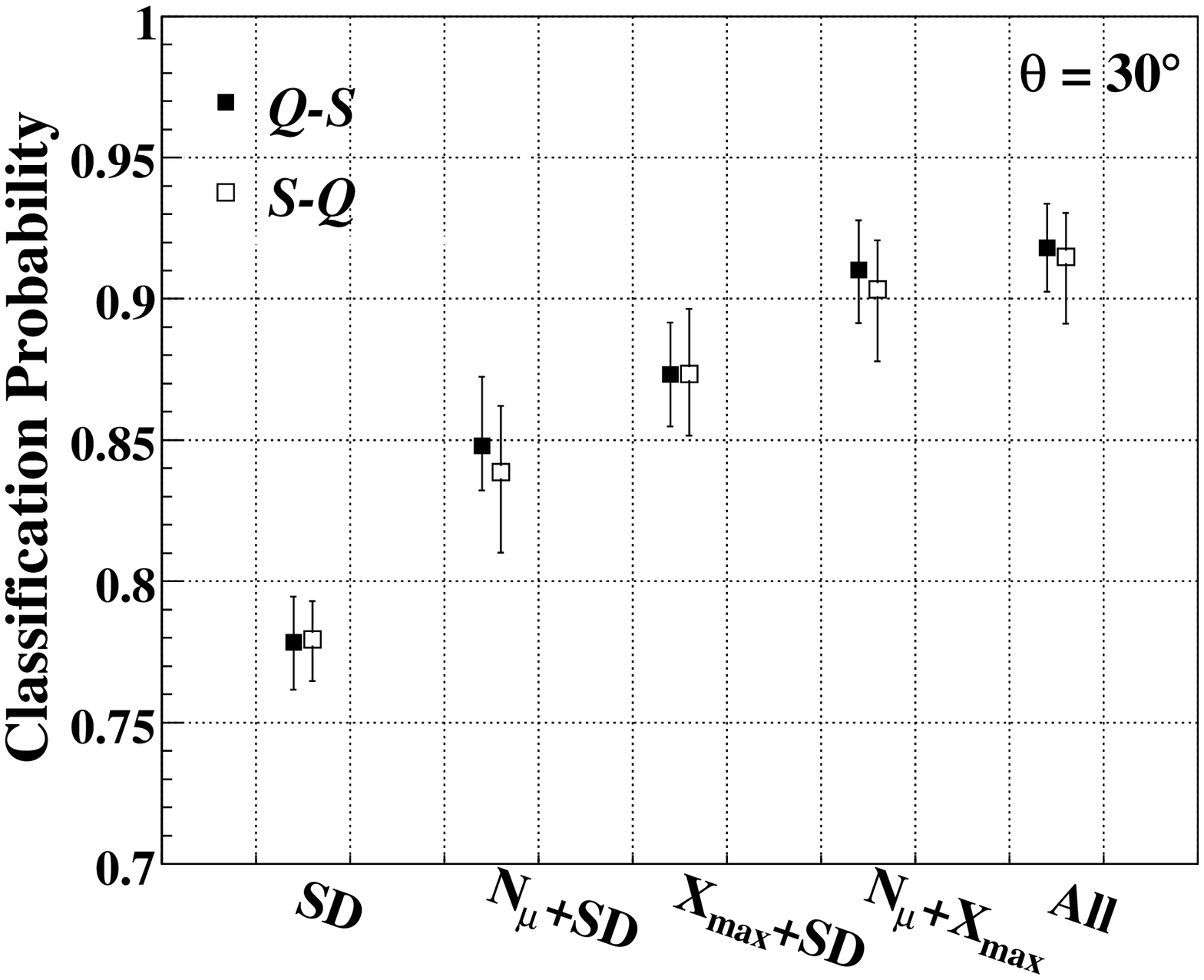}
\includegraphics[width=6.8cm]{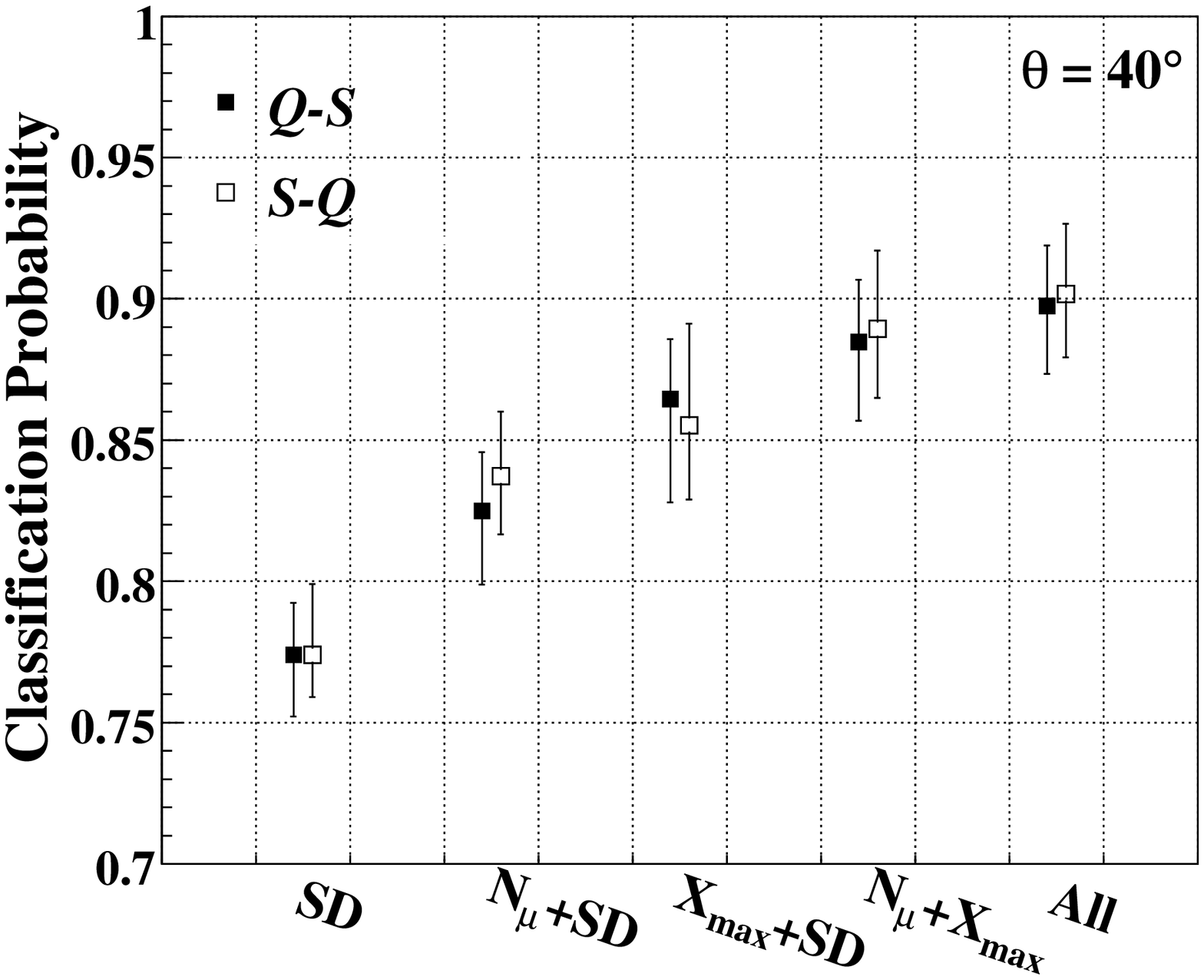}
\caption{Classification probability (medians and regions of $68\%$ of probability) for the different 
sets of parameters considered, $\theta = 30^\circ$ and $\theta = 40^\circ$ assuming that QGSJET-II is the 
true hadronic interaction model whereas the real one is Sibyll 2.1 ($Q-S$) and vice versa ($S-Q$).
\label{PclMet2}}
\end{center}
\end{figure}

Figure \ref{PclMet2} shows the medians and the regions of $68\%$ of probability of the classification 
probability distributions for the different sets of parameters considered, $\theta = 30^\circ$ and 
$\theta = 40^\circ$ and for both cases, $Q-S$ and $S-Q$. The classification probabilities, 
for the different combinations of parameters considered and for both zenith angles, result compatibles 
between $Q-S$ and $S-Q$ cases. This happens because the hadronic interaction models considered produce
observable parameters that are quite similar between themselves. Besides, the values of the classification 
probabilities for the cases, $Q-S$ and $S-Q$, are compatibles with the ones obtained considering each one 
of those hadronic interaction models used to obtain both, the density estimates and test samples.  

Note that, the previous analyses are done by using samples of events which are not independent 
because each shower is used 50 times to simulate the response of the detectors (see, however, Ref. 
\cite{Repetitions:08}). This fact does not affect the present calculation, see appendix \ref{NonInd} 
for details.

We center here in the Auger enhancements. Consequently, we particularize 
our analysis for the specific properties of these detectors. Regarding detector resolution, 
we assume an energy error of $25$\%, which is much larger than the one obtained at present by 
the current Auger calibration with hybrid events ($\sim 18$\%). In the same way, the error in
$X_{max}$,  $\Delta X_{max}$, is consistent with those obtained by the Auger fluorescence 
detectors ($\sim 20$ g cm$^{-2}$ at $\sim 1$ EeV). The resolution on the determination of 
$N_{\mu}(600)$ ($< 12$\% for $E\geq1$ EeV) has been calculated in Ref. \cite{SupaRec:08} 
specifically for the muon counters under consideration. Experimental uncertainties related 
with the simulation of the Cherenkov detectors or the reconstructed position of the core 
have been discussed extensively in Ref. \cite{Ghia:07} and \cite{Medina:06}, respectively, 
and their results have been incorporated in the calculation of the discrimination probabilities 
presented here.

Nevertheless, the muon counters are the only component of the system that
has not been built yet beyond the prototype stage and, therefore, might be
thought of as more uncertain. It can be shown that, by adding Gaussian fluctuations 
of magnitude $\alpha$, from an unspecified origin and applied to each muon counter, the 
relative error in the determination of $N_{\mu}(600)$ increases less than $5\%$
for $\alpha \lesssim 25\%$, which would increase the relative error from $12\%$ to 
$17\%$. At this level, the error in energy, already included in our model, 
would still be dominant, leaving our conclusions unchanged.

\subsection{Zenith angle dependence}
\label{Zenith}

The distribution functions of the different combinations of mass sensitive parameters depend on zenith angle. 
Therefore, as shown in figure \ref{PclMet1}, the classification probability also depends on it. 

The parameter $X_{max}$ weakly depends on $\theta$ but $N_{\mu}(600)$ do not. The simulations show that
a new parameter, based on $N_{\mu}(600)$, but weakly dependent on zenith angle is obtained from,
\begin{equation}
\label{NmuTilde}
\tilde{N}_{\mu}(600)=N_{\mu}(600) \left[ \frac{\cos\theta_{ref}}{\cos\theta} \right]^{3/2}.
\end{equation}
where $\theta_{ref}=30^\circ$ is a reference angle. In this way a pair of mass sensitive parameters, 
$\{\tilde{N}_{\mu}(600),\ X_{max} \}$, that are almost constant with $\theta$ is obtained. Therefore, the 
estimates corresponding to $\theta=30^\circ$ can be used to obtain the classification probability as a function 
of zenith angle for this pair of parameters.

The 20 pairs of proton and iron samples, corresponding to $\theta=30^\circ$ and QGSJET-II, are used to construct 
the corresponding density estimates for $\vec{x}=(\tilde{N}_{\mu}(600),X_{max})$. Then the single pairs of 
proton and iron samples corresponding to $\theta=20^\circ,\ 25^\circ,\ 35^\circ,\ 40^\circ,\ 45^\circ$ (just 
one pair of proton and iron samples of the 20 available for $\theta=40^\circ$ is considered) is used to calculate 
the classification probability as a function of $\theta$, see appendix \ref{PclVsZhetaApp} for details of the 
calculation.

\begin{figure}[!bt]
\begin{center}
\includegraphics[width=12cm]{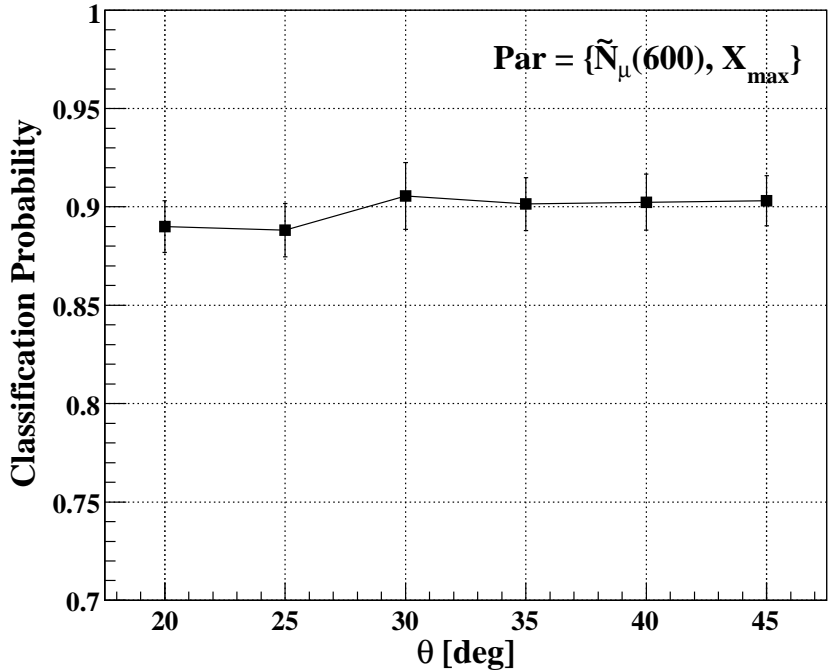}
\caption{Classification probability as a function of zenith angle corresponding to $\{\tilde{N}_{\mu}(600),\ X_{max} \}$
and hadronic interaction model QGSJET-II. The proton and iron nuclei density estimates of this pair of parameters used 
for the analysis correspond to QGSJET-II and $\theta= 30^\circ$ (see text for details).\label{PclVsTheta}}
\end{center}
\end{figure}

Figure \ref{PclVsTheta} shows the obtained classification probability as a function $\theta$. The result corresponding 
to $\theta= 30^\circ$, obtained in the previous subsection (see table \ref{tpclqq30}), is included in the figure. 
From this figure it can be seen that the classification probability is almost constant with $\theta$ and the mean value 
is approximately $0.9$.  

\section{Composition of a sample}
\label{compo}

The determination of the composition of a sample of size $N=N_{pr}+N_{fe}$, where $N_{pr}$ and $N_{fe}$ are 
the number of protons and iron nuclei, respectively, consists in the estimation of the proton content, i.e., 
$c_{p}=N_{pr}/N$. For that purpose suppose that the classification probabilities obtained for a given set of 
parameters are $P_{cl}^{pr}$ and $P_{cl}^{fe}$ for protons and iron nuclei, respectively. Therefore, the number 
of proton and iron events correctly classified follow the binomial distribution,
\begin{eqnarray}
\label{BinPr}
P(n_{pr}) &=& {N_{pr} \choose n_{pr}}\ (P_{cl}^{pr})^{N_{pr}-n_{pr}}\ (1-P_{cl}^{pr})^{n_{pr}}, \\
\label{BinFe}
P(n_{fe}) &=& {N_{fe} \choose n_{fe}}\ (P_{cl}^{fe})^{N_{fe}-n_{fe}}\ (1-P_{cl}^{fe})^{n_{fe}}.
\end{eqnarray}

The number of events classified as protons or as iron nuclei are $n_{pr}^{cl} = n_{pr}+N_{fe}-n_{fe}$ and
$n_{fe}^{cl} = n_{fe}+N_{pr}-n_{pr}$, respectively. By using Eqs. (\ref{BinPr}, \ref{BinFe}) to calculate the 
expected values of these quantities the following expressions are obtained,
\begin{equation}
\label{Pmatrix}
\left(
\begin{array}{c}
  E[n_{pr}^{cl}] \\
  E[n_{fe}^{cl}] \\
\end{array}
\right) =
\mathbf{P}
\left(
\begin{array}{c}
  N_{pr} \\
  N_{fe} \\
\end{array}
\right) =
\left(
\begin{array}{cc}
  P_{cl}^{pr}   & 1-P_{cl}^{fe}\\
  1-P_{cl}^{pr} & P_{cl}^{fe} \\
\end{array}
\right)
\left(
\begin{array}{c}
  N_{pr} \\
  N_{fe} \\
\end{array}
\right).
\end{equation}
Last equation suggests the following definition of estimators of $N_{pr}$ and $N_{fe}$ \cite{Antoni:07},
\begin{equation}
\label{Est}
\left(
\begin{array}{c}
  \hat{N}_{pr} \\
  \hat{N}_{fe} \\
\end{array}
\right) = \mathbf{P}^{-1}
\left(
\begin{array}{c}
  n_{pr}^{cl} \\
  n_{fe}^{cl} \\
\end{array}
\right),
\end{equation}
which, as a function of $n_{pr}$ and $n_{fe}$ give,
\begin{eqnarray}
\label{NpEst}
\hat{N}_{pr} &=& \frac{N_{fe} P_{cl}^{fe}+N_{pr} (P_{cl}^{fe}-1)+n_{pr}-n_{fe}}{P_{cl}^{pr}+P_{cl}^{fe}-1}, \\ 
\label{NfEst}
\hat{N}_{fe} &=& \frac{N_{pr} P_{cl}^{pr}+N_{fe} (P_{cl}^{pr}-1)+n_{fe}-n_{pr}}{P_{cl}^{pr}+P_{cl}^{fe}-1}.
\end{eqnarray}

Note that, by construction, $E[\hat{N}_{pr}] = N_{pr}$ and $E[\hat{N}_{fe}] = N_{fe}$ (the estimates are non-biased).
Therefore, the composition estimator $\hat{c}_p = \hat{N}_{pr}/N$ is also non-biased and the variance, obtained from Eq. 
(\ref{NpEst}) using (\ref{BinPr}) and (\ref{BinFe}), is
\begin{equation}
\label{SigCp}
Var[\hat{c}_p] = \frac{1}{N}\ \frac{P_{cl}^{pr} (1-P_{cl}^{pr}) c_p+P_{cl}^{fe} (1-P_{fe}^{pr}) (1-c_p)}%
{(P_{cl}^{pr}+P_{cl}^{fe}-1)^2}.
\end{equation}
Note that the variance is inversely proportional to the sample size.

As shown in subsection \ref{MethRes}, $P_{cl}^{pr}$ and $P_{cl}^{fe}$ are random variable. This happens because finite 
samples of events are used as test samples as well as to construct the density estimates. Therefore, the distributions 
of $P_{cl}^{pr}$ and $P_{cl}^{fe}$, obtained in subsection \ref{MethRes}, correspond to samples of $\sim 2500$ events 
to construct the density estimates as well as for the test samples size.

The lowest values corresponding to the region of $68\%$ of probability of the distributions of $P_{cl}^{pr}$ 
and $P_{cl}^{fe}$ (see table \ref{tpclqq30} for $\theta = 30^\circ$ and QGSJET-II) are used to obtain an upper limit of 
the uncertainty in the determination of the composition of a sample, $\sigma[\hat{c}_p]=Var[\hat{c}_p]^{1/2}$. This 
approximation overestimates the error because, as mentioned, the classification probability distributions also 
include the fluctuations due to the finite size of the test samples. In fact, the standard deviation for these fluctuations 
is $\sigma_{fs}=[p\, (1-p)/N]^{1/2}$, where $p$ is the classification probability of protons or iron nuclei corresponding 
to a given pair of density estimates. $\sigma_{fs} \lesssim 0.01$ for samples of $\sim 2500$ events and $p=0.7-0.9$. From table 
\ref{tpclqq30}, it can be seen that such fluctuations are comparable to the total fluctuations. Test samples of much larger 
size are needed to obtain the classification probability distributions corresponding to density estimates built from samples 
of $\sim 2500$ events, without including such fluctuations. Note that also changing the number of events of the samples used 
to construct the density estimates would modify the fluctuations on the classification probability distributions, in fact, such 
fluctuations can be reduced by using samples of larger number of events. 
 
Figure \ref{SigCp} shows $\sigma[\hat{c}_p]$ as a function of $c_p$ for the different sets of parameters considered, 
$\theta = 30^\circ$, QGSJET-II and for samples of 100 events, which is the number of hybrid events expected for the energy 
interval $\Pi_r$ in two years of data taking of AMIGA and HEAT \cite{Medina:06}. As mentioned, $\sigma[\hat{c}_p]$ is calculated 
by using the lowest values of the region of $68\%$ of probability corresponding to the proton and iron classification probability 
distributions. From this figure it can be seen that $\sigma[\hat{c}_p]$ varies very slowly with $c_p$ because the  
classification probability for protons and iron nuclei are quite similar (see table \ref{tpclqq30}). As expected, the larger 
the proton and iron classification probabilities the smaller the values of $\sigma[\hat{c}_p]$.    
\begin{figure}[!bt]
\begin{center}
\includegraphics[width=12cm]{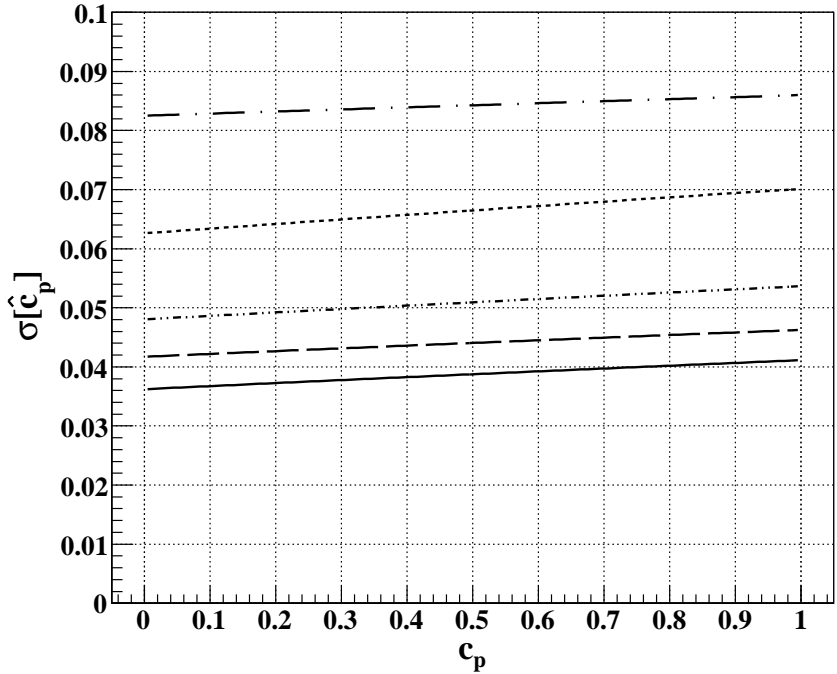}
\caption{$\sigma[\hat{c}_p]$ as a function of $c_p$ obtained by using the lowest values of region of $68\%$ of probability 
corresponding to the proton and iron classification probability distributions, for samples of 100 events, $\theta = 30^\circ$ 
and hadronic interaction model QGSJET-II. The curves correspond to: from bottom to top, $All$ parameters, $\{N_\mu(600), X_{max}\}$, 
$\{X_{max}, SD \}$, $\{N_\mu(600), SD\}$ and $SD$.\label{SigCp}}
\end{center}
\end{figure}

\section{Conclusions}
\label{conc}

Under the assumption of a binary mixture of proton and iron nuclei, we study the possibility 
of assigning a statistically meaningful classification probability to individual cosmic 
ray showers in the ankle region of the spectrum. We use a non-parametric technique to 
estimate the relevant multi-dimensional density functions for the different combinations of 
parameters that will be available from the Auger enhancements AMIGA and HEAT. 

We find that, as expected, the maximum classification probability is obtained by combining 
simultaneously all surface and fluorescence parameters ($\sim 92\%$ for QGSJET-II). However, 
the two leading parameters are $\{N_\mu(600), X_{max}\}$ which, used in a two-dimensional analysis, 
already give a classification probability almost as large as the whole set of parameters combined 
($\sim 91\%$ for QGSJET-II). This means that, the separation between protons and iron nuclei, is 
mainly expressed experimentally through these two parameters.

We also study the effect of assuming QGSJET-II as the true hadronic interaction model (used to construct 
the density estimates) when the real one is Sibyll 2.1 (used to obtain the test samples) and vice versa. 
We find that the classification probabilities obtained in either case, have systematic differences which 
are much smaller than their statistical fluctuations. This happens because the hadronic interaction models 
considered produce similar air showers and, consequently, similar mass sensitive parameters. Nevertheless, 
other models, like EPOS \cite{EPOS:06}, which predict a very different shower muon content, would affect the 
composition estimates obtained with our technique. If the later were the case, a possible way out might come 
from the application of a bi-dimensional technique, in which $X_{max}$ and $N_{\mu}$ information from hybrid 
events is used simultaneously in order to disentangle the hadronic uncertainty from the composition effects 
(see Ref. \cite{SupaCompo:08}).   

Although the values of the classification probability, obtained for the different sets of parameters 
considered, do not allow a reliable classification into proton and iron primaries on an event-by-event bases, 
the composition of a sample can be estimated with reasonable accuracy using this technique. It must be kept in 
mind, however, that any composition estimate inside this framework, is bounded by the simplifying assumptions of 
a binary mixture and of a hadronic interaction model that is well represented by either QGSJET-II or Sibyll 2.1. 
The limitation to a mixture of two components is not actually a limitation of our mathematical formulation, as 
expressed for example in Section \ref{clastech}, but a realization of the limitations posed by intrinsic shower 
fluctuations, detector uncertainties and limitations and the lack of a precise knowledge of the hadronic 
interactions. Regarding the latter, the most widely used models in the literature are at present QGSJET and Sibyll 
in their latest versions, which we use in the present work. Even if the differences between these two models are 
not large enough as to render our results ambiguous, these models could be wrong. In fact, as already mentioned, 
the new model EPOS predicts a much higher number of muons for a given nuclei at any energy. Needless to say, these 
constraints apply not only to our own, but to any technique that attempts to determine composition, and they are 
important factors to which accelerator particle physics will certainly make a fundamental contribution in the near 
future.

\section{Acknowledgments}

The authors have greatly benefited from discussions with several colleagues
from the Pierre Auger Collaboration, of which they are members. GMT acknowledges 
the support of DGAPA-UNAM through grant IN115707.

\appendix

\section{Effect of the non independence of the events}
\label{NonInd}

The computer processing time and disk space required to obtain several samples of showers with good statistics are very 
large. Therefore, each simulated shower is used $50$ times, to increase the statistics, by uniformly distributing impact 
points in the array area (see section \ref{sim}). The simulated events generated in this way, corresponding to a given 
sample, are not independent. 

A pair of proton and iron samples of 1000 independent events each (showers after detectors simulations) are obtained 
by taking at random 50 independent events corresponding to 50 independent showers of each of the 20 proton and 20 iron 
samples available for each zenith angle ($30^\circ$ and $40^\circ$) and hadronic interaction model considered.
Then, the leave-one-out technique \cite{Chilingarian:89} is used to estimate the classification probabilities for each 
pair of proton and iron samples and combination of parameters considered.

Table \ref{tpclqq30is} shows the classification probabilities obtained for $\theta = 30^\circ$ and QGSJET-II for the different
combination of parameters considered. Although the number of events corresponding to the proton and iron samples composed by 
independent events is $\sim 2.4$ times smaller than the one used for the analysis of subsection \ref{MethRes}, the obtained 
values of the classification probabilities $P_{cl}^{pr}$, $P_{cl}^{fe}$ and $P_{cl}$ of table \ref{tpclqq30is} are contained 
in the region of $68\%$ of probability of the corresponding distributions for the non-independent event samples (see table 
\ref{tpclqq30}).   
\begin{table}[h]
\begin{center}
\caption{Classification probability corresponding to protons, iron nuclei and both together, obtained from two samples,
one for protons an the other for iron, of 1000 independent events each, for $\theta = 30^\circ$ and high energy hadronic
interaction model QGSJET-II.}
\label{tpclqq30is}
\begin{tabular}{c c c c} \hline
Parameters      & $\ \ P_{cl}^{pr}\ \ $ &  $\ \ P_{cl}^{fe}\ \ $ &  $\ \ P_{cl}\ \ $     \\  \hline 
$All$           &      $0.88$           &        $0.92$          &       $0.90$          \\ 
$N_\mu+X_{max}$ &      $0.88$           &        $0.91$          &       $0.89$          \\ 
$X_{max}+SD$    &      $0.84$           &        $0.88$          &       $0.86$          \\ 
$N_\mu+SD$      &      $0.83$           &        $0.84$          &       $0.83$          \\ 
$SD$            &      $0.74$           &        $0.79$          &       $0.77$          \\  \hline 
\end{tabular}
\end{center}
\end{table}

Similar results were obtained for $\theta=40^\circ$ and for Sibyll 2.1 as well as for the cases in which one hadronic 
interaction model is used to construct the density estimates and the other for the test samples. For all these cases
the results obtained for the samples composed by independent events are compatible with the corresponding to the 
samples composed by non-independent events.

\section{Calculation of the classification probability using a single test sample}
\label{PclVsZhetaApp}

In this appendix details of the calculation of the classification probability and their uncertainties corresponding to 
the pair of parameters $\{\tilde{N}_{\mu}(600),\ X_{max} \}$ (see subsection \ref{Zenith}) are given. 

The pair of proton and iron samples of number of events $N_{pr}$ and $N_{fe}$, respectively, for a given 
zenith angle is used together with 20 pairs of proton and iron density estimates corresponding to $\theta = 30^\circ$. 
In this way, 20 different values for protons and 20 for iron nuclei of the number of events correctly classified are 
obtained (see Eqs. (\ref{pclpr}, \ref{pclfe}, \ref{pcl})): $n_{pr}^0(s)$ and $n_{fe}^0(s)$ with $s=1,\ldots,M$ and 
$M=20$.   

The distribution function of the number of events correctly classified is binomial, therefore, including the uncertainty
in the determination of the parameter $p$ of the binomial distribution, inferred from a given number of positive trials 
$n_{A}^0(s)$, the following expression is obtained,
\begin{equation}
\label{pna}
P(n_A | N_A, n_{A}^0(s)) = {N_{A} \choose n_{A}}\ \int_0^1 dp\ p^{n_A} (1-p)^{N_A-n_A} B(p|n_{A}^0(s)+1,N_A+n_{A}^0(s)-1),
\end{equation} 
where, 
\begin{equation}
B(p|\alpha,\beta) = \frac{\Gamma(\alpha+\beta)}{\Gamma(\alpha)\ \Gamma(\beta)}\ p^{\alpha-1} (1-p)^{\beta-1},
\end{equation}
is the Beta distribution and $\Gamma(x)$ is the Gamma function.

The distribution function of $n_A$ can be estimated from,
\begin{equation}
f(n_A)=\frac{1}{M}\ \sum_{s=1}^{M} P(n_A | N_A, n_{A}^0(s)).
\end{equation}

Therefore, the mean value and standard deviation of the classification probability, $P_{cl}^{A}=n_A/N_A$, are
given by,
\begin{eqnarray}
E[P_{cl}^{A}] &=& \frac{1}{M N_A}\ \sum_{s=1}^{M} E[n_A(s)], \\
\sigma[P_{cl}^A]&=& \frac{1}{N_A}\ \left[ \frac{1}{M} \sum_{s=1}^M (\sigma[n_{A}(s)]^2 +% 
E[n_{A}(s)]^2) -E[P_{cl}^{A}]^2 \right]^{1/2},
\end{eqnarray}
by using Eq. (\ref{pna}) they become,
\begin{eqnarray}
E[n_A(s)] &=& N_A \ \frac{n_{A}^0(s)+1}{N_A+1}, \\
\sigma[n_{A}(s)]^2 &=& E[n_A(s)] (1-E[n_A(s)]) + \nonumber \\
&& \frac{(N_A-1)(N_A-n_{A}^0(s)+1)(n_{A}^0(s)+1)}{(N_A+2)^2 (N_A+3)}.
\end{eqnarray}
 
Finally, the mean value and the standard deviation of $P_{cl}$ are estimated from, 
\begin{eqnarray}
E[P_{cl}] &=& \frac{1}{M N}\ \sum_{s=1}^{M} E[n(s)], \\
\sigma[P_{cl}]&=& \frac{1}{N}\ \left[ \frac{1}{M} \sum_{s=1}^M (\sigma[n(s)]^2 +% 
E[n(s)]^2) -E[P_{cl}]^2 \right]^{1/2},
\end{eqnarray}
where $N=N_{pr}+N_{fe}$, $E[n(s)] = E[n_{pr}(s)]+E[n_{fe}(s)]$ and 
$\sigma[n(s)]^2=\sigma[n_{pr}(s)]^2+\sigma[n_{pr}(s)]^2$.

\end{document}